\documentclass{gHPR2e}

\usepackage{subfigure}

\begin{document}



\title{High pressure research using muons at the Paul Scherrer Institute}

\author{R. Khasanov$^{\rm a}$$^{\ast}$\thanks{$^\ast$Corresponding author. Email: rustem.khasanov@psi.ch
\vspace{6pt}},
Z. Guguchia$^{\rm a}$,
A.~Maisuradze$^{\rm a}$,
D.~Andreica$^{\rm b}$,
M.~Elender$^{\rm a}$,
A.~Raselli$^{\rm a}$,
Z.~Shermadini$^{\rm a}$,
T.~Goko$^{\rm a}$,
F.~Knecht$^{\rm a}$,
E.~Morenzoni$^{\rm a}$,
and A.~Amato$^{\rm a}$
\\\vspace{6pt}  $^{a}${\em{Laboratory for Muon Spin Spectroscopy, Paul Scherrer Institut, CH-5232 Villigen PSI, Switzerland}}\\
$^{b}${\em{Faculty of Physics, Babes-Bolyai University, 400084 Cluj-Napoca, Romania}}\\\received{XXXX 2016} }

\maketitle

\begin{abstract}
Pressure, together with temperature and magnetic field, is an important thermodynamical parameter in physics. Investigating the response of a compound or of a material to pressure allows
to elucidate ground states, investigate their interplay and interactions and determine microscopic parameters.  Pressure tuning is used to establish phase diagrams, study phase transitions and identify critical points.
Muon spin rotation/relaxation ($\mu$SR) is now a standard technique making increasing significant contribution in condensed matter physics, material science research and other fields.
In this review, we will discuss specific requirements and challenges to perform $\mu$SR experiments under pressure, introduce the high-pressure muon facility at the Paul Scherrer Institute (PSI, Switzerland) and present selected results obtained by combining the sensitivity of the $\mu$SR technique with pressure.


\end{abstract}

\section{Introduction}

Muon spin rotation/relaxation ($\mu$SR) experiments make use of polarized muons, which thermalized in the sample, act as a magnetic spin microprobe or as a hydrogen-like probe in the host material. Muons provide unique information at local atomic level on physical properties of matter and are used to study a variety of static and dynamic phenomena in superconductivity, magnetism, material science, radical chemistry, semiconductor physics and many other fields. They give information unavailable to other techniques so that muon studies are often complementing those involving other methods.
High pressure research has become one of the most significant area in $\mu$SR and is a major tool in understanding the properties of different phases of matter and their interplay. This is best exemplified in the role of magnetism in many unconventional superconductors, where pressure can drive the system from one state to another and where the specific local information of $\mu$SR can help to identify on one hand different magnetic states (ordered, short range ordered or disordered) and how they coexist or compete with superconductivity at nanoscale and, on the other hand, to quantify  the effects on the strength and structure of superconductivity, such as the determining the superfluid density and the gap symmetry and structure.
The importance of this technique has led to the development of dedicated instruments. The instrument called GPD (General Purpose Decay instrument) and the associated beam-line at PSI represent the state of the art of the tools available for pressure studies in $\mu$SR; to enhance the field of application a continuous effort is being invested in improving the relevant parameters for an experiment such as pressure and temperature range, signal to background level and minimum sample quantity required.

In this review paper we summarize the status of high-presssure studies in $\mu$SR. The paper is organized as follows.  Section \ref{Sec:muon-technique} gives a short introduction of the $\mu$SR method. The experimental setup at PSI  with the $\mu$E1 beam line, delivering the energetic positive muons to the GPD spectrometer are described in Sec.~\ref{Sec:muSR_setup}.
The specificities of a $\mu$SR experiment under pressure are pointed out in this Section ~\ref{Sec:muSR_under_pressure}. In Section ~\ref{sec:Pressure_cell} details of the design of the piston-cylinder pressure cells
which are used to apply pressure up to $\sim$ 2.4 GPa to the sample are given. 
Material considerations and the pressure calibration procedure at low temperature are presented as well. Some scientific examples showing the capabilities of $\mu$SR under pressure for studying a large variety of phenomena are discussed in Section \ref{Sec:Examples}.
The paper concludes with a brief outlook about future possible directions of development.

\section{Introduction to the $\mu$SR technique} \label{Sec:muon-technique}

In this section we give a brief introduction to the $\mu$SR technique. The interested reader is referred to more comprehensive textbooks \cite{Yaouanc_BOOK_2011,Schenck_BOOK_1985}.
The muon is an elementary particle, positively or negatively charged, with spin 1/2 and lifetime  $\tau_\mu\simeq2.2$~$\mu$s. $\mu$SR makes mostly use of the positive muons, which can be produced with very high polarization. Implanted in a solid sample they very quickly thermalize at a generally interstitial position of the lattice where they act as a local magnetic probe.
The possibility to use  muons as microprobe of matter was predicted already in 1957 in the seminal work of L.R.~Garwin et al. reporting about ''Observations of the Failure of Conservation of Parity and Charge Conjugation in Meson Decays: the Magnetic Moment of the Free Muon`` \cite{Garwin_PR_57}, where they recognized the potential of polarized muons and wrote: ''It seems possible that polarized positive and negative muons will become a powerful tool for exploring magnetic fields in nuclei, atoms, and interatomic regions``. Muons are obtained from the decay of positive pions ($\pi^{+} \rightarrow \mu^{+} + \nu_{\mu}$). At PSI the pions are generated by bombarding two graphite targets with an intense beam of protons with 590 MeV kinetic energy at a present maximum current of 2.2 mA. Because of the parity violating pion decay only left-handed $\nu_{\mu}$ exist so that the decay muons have the spin antiparallel to their momentum in the pion rest frame. This property allows to obtaining polarized beams.
Presently, there are worldwide four facilities: PSI (Switzerland) and TRIUMF (Canada), delivering continuous beams and ISIS (UK) and J-PARC (Japan), delivering pulsed beams. In these facilities beam lines have been optimized to transport polarized muons to dedicated instruments where $\mu$SR experiments under different experimental conditions are performed.

The $\mu$SR technique is based on the observation of the time evolution of the muon spin polarization $P(t)$ of an ensemble of muons after their implantation into a sample ($t=0$).
The direct observation of the muon spin dynamics is possible due to the parity violating muon decay. Positive muons decay into a positron, muon antineutrino and electron neutrino ($\mu^{+} \rightarrow e^{+} + \nu_{e}+ \bar{\nu}_{\mu}$). The positron is preferentially emitted along the direction of the muon spin at a moment of the decay (decay asymmetry). Thus, by measuring the direction and the timing of a significant number of decay positrons (typically several millions per detector) it is possible to directly follow the time evolution of the polarization of a muon ensemble.

\begin{figure}[t!]
\centering
\includegraphics[width=0.9\linewidth]{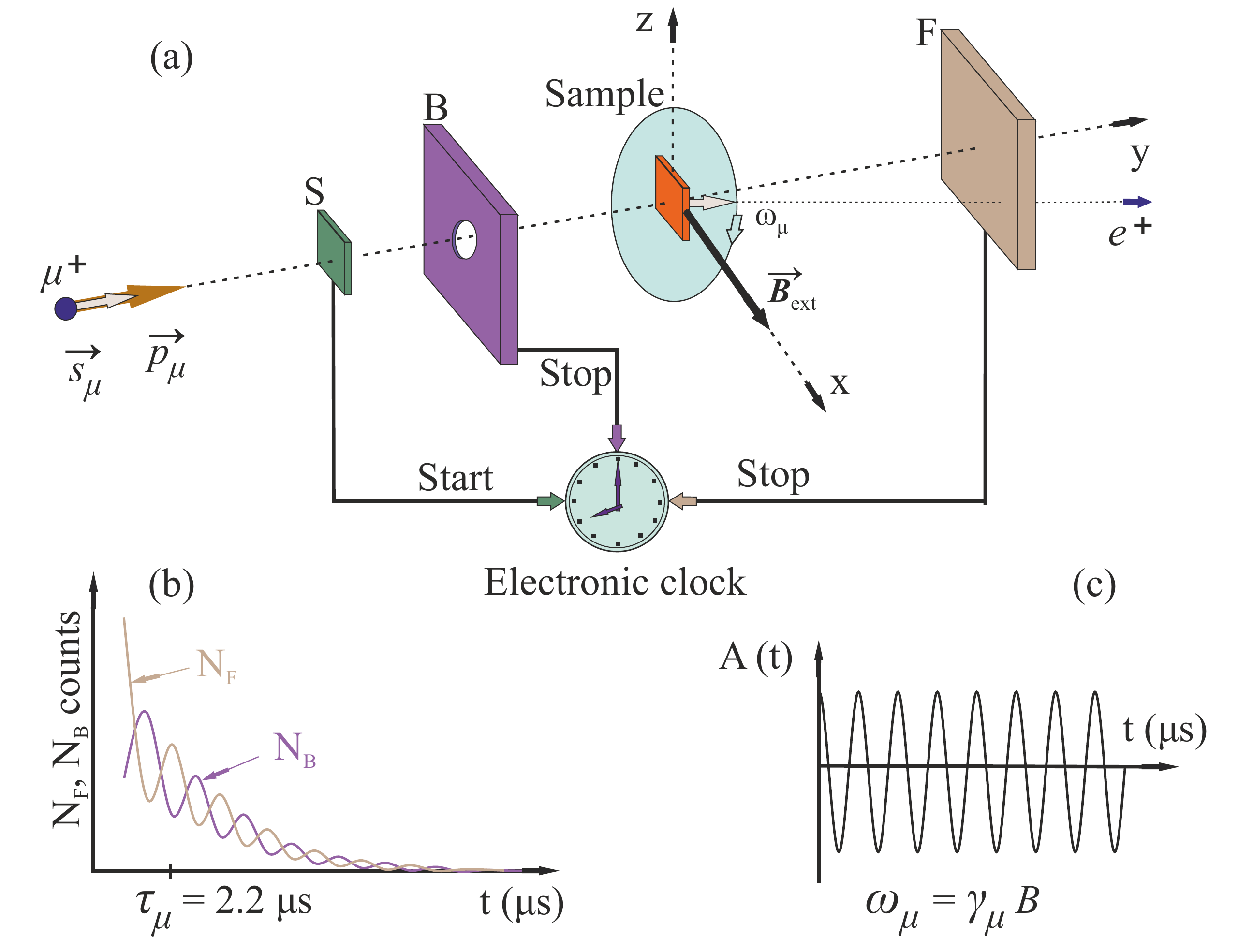}
\vspace{0cm}
\caption{\footnotesize Principle of a ${\mu}$SR experiment. (a) overview of the experimental setup.
Spin polarized muons with spin $\vec{s}_{\mu}$ parallel to the momentum $\vec{p}_{\mu}$ are implanted  in the sample placed between the forward (F) and the backward (B) positron detectors. A clock is started when the incoming muon traverses the start detector (S) and stopped by the decay positron detected in detector F or B. (b) Histograms of detected positrons $N_{F}$ and $N_{B}$ as a function of time for the forward and backward detector, respectively. (c) The asymmetry signal calculated according to Eq.~(\ref{eq_muSR_asymmetry}).}
\label{Fig:MSR_schematic}
\end{figure}

The basic principle of a $\mu$SR experiment at the high pressure setup at PSI is illustrated in Fig.~\ref{Fig:MSR_schematic}.
At the time of the implantation into the sample, a clock triggered by the muon traversing detector S is started. After the very quick thermalization, the muons spin starts to precess with the Larmor frequency $\omega_\mu=\gamma_\mu\cdot B_{\rm loc}$ ($\gamma_\mu=2\pi\cdot135.5$\,MHz/T is the gyromagnetic ratio of the muon) in the local magnetic field $B_{\rm loc}$ ($\vec{B}_{\rm loc}$vector sum of external and internal magnetic field) until it decays with lifetime of $\tau_\mu\simeq2.2\,\mu$s and emits a positron preferentially in the direction of the spin at the time of decay.
The latter is then detected by one of the positron-detectors which stops the clock.
Repeating this procedure for several $10^6$ muons, histograms ($\mu$SR time spectra) are generated.
For the initial spin orientation of the muon parallel to its momentum (as is the case for the energetic muons used for pressure studies), the spin rotation/relaxation can be observed using two positron counters mounted on the opposite sides of the sample in the forward $N_{\rm F}(t)$ and the backward $N_{\rm B}(t)$ detector(F/B denotes forward/backward with respect to the initial spin direction).
The number of positrons detected by each counter as a function of time reflects the time dependence of the muon spin polarisation along the axis of observation $\hat{n}$ defined by the two detectors:
\begin{equation}
N_{\rm F(B)} (t)=N_0 \exp[-t/\tau_\mu]\cdot(1 \pm A_{\rm 0}P(t){N}_{\rm F(B)}),
\end{equation}
where
\begin{equation}
 P(t)=\frac{\left< \vec{s}_{\mu}(t)\cdot \hat{n} \right>}{\left|\vec{s}_{\mu}(0)\right|}
\end{equation}
$A_{\rm 0}$ is the experimentally observable maximum decayasymmetry at $t=0$ and the average is over the muon ensemble implanted in the sample. Its value depends on the details of the experimental configuration and factors, such as the detector solid angle and efficiency, absorption and scattering of positrons in the material. It typically lies between 0.25 and 0.3, slightly lower than the intrinsic asymmetry of the muon decay, which is 1/3. The signal corresponding to the time evolution of the muon spin polarisation can be directly extracted by looking at the so called asymmetry function $A(t)$, determined as the difference of the signals observed by the two counters, normalised by their sum:
\begin{equation}
 A(t)=A_{0}P(t)=\frac{N_{\rm F}(t)-\alpha N_{\rm B}(t)}{N_{\rm F}(t)+\alpha N_{\rm B}(t)}
\label{eq_muSR_asymmetry}
\end{equation}
The parameter $\alpha$ takes into account the different solid angle and efficiency of the F and B positron-detectors and is calibrated at the beginning of the experiment.
The quantities $A(t)$ or $P(t)$ contain all the information about the interaction of the muon spin (or magnetic moment) with its local environment and provide therefore the physical information about the investigated system.

\begin{figure}[htb]
\centering
\includegraphics[width=0.5\linewidth]{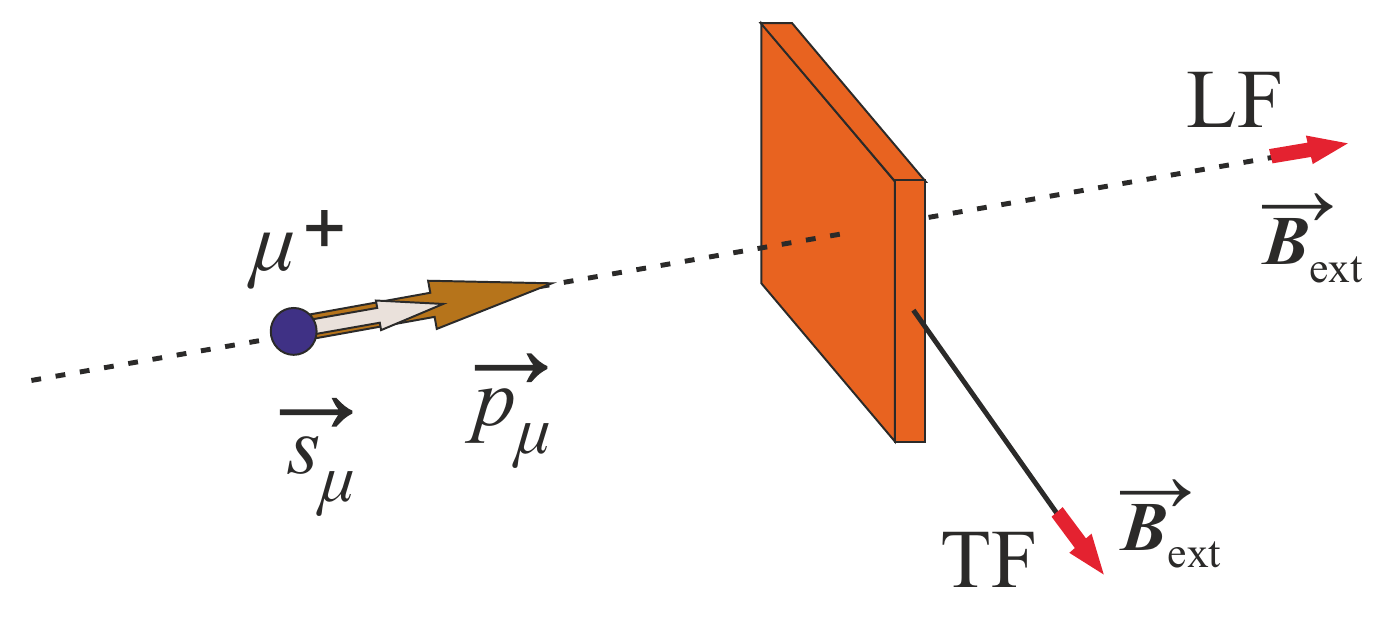}
\caption{\footnotesize Schematics of TF- and a LF-${\mu}$SR configurations (see text for explanation). }
\label{Fig:TF-LF}
\end{figure}

Generally two different magnetic field configurations are used: the transverse-field (TF) and  the longitudinal (LF) configuration. In the TF-configuration the external magnetic field $\vec{B}_{\rm ext}$ is applied perpendicular to the initial muon polarization, whereas in LF-$\mu$SR the field is applied along the initial muon spin direction (see Fig.~\ref{Fig:TF-LF}). The $\mu$SR experiment can be also performed in zero applied field (ZF).

\section{Setup for $\mu$SR pressure studies} \label{Sec:muSR_setup}

In this section we describe the $\mu$E1 beam line, delivering high energetic polarized positive muons, and the GPD instrument.

\subsection{The $\mu$E1 beam line} \label{Sec:mue1}

In conventional $\mu$SR experiments in bulk materials the so-called surface muons are used. They are obtained from pions decaying at rest near the surface of the production target and hence their name. As a consequence of the two-body decay they are fully polarized and have a kinetic energy of 4.1 MeV and a momentum of 29.8 MeV/c, so that beams based on this process are nearly monochromatic and 100\% polarized. Their range in matter corresponds to $\sim 150-200$ mg/cm$^2$ ({\it e.g.} in Cu with density 8.96~g/cm$^3$ this amounts to 0.2--0.3~mm only). To be able to traverse the cell walls and stop in the sample material placed inside a pressure cell, more energetic muons are needed. This is achieved by collecting muons, which are the product of pions decaying in flight, so that the total momentum of the muons is the relativistic vector sum of the pion momentum and of the intrinsic momentum from the decay. The polarization of such a high energy muon beam (decay muon beam line) in the laboratory frame is about 80\%.

\begin{figure}[htb]
\centering
 \includegraphics[width=0.7\linewidth, angle=0]{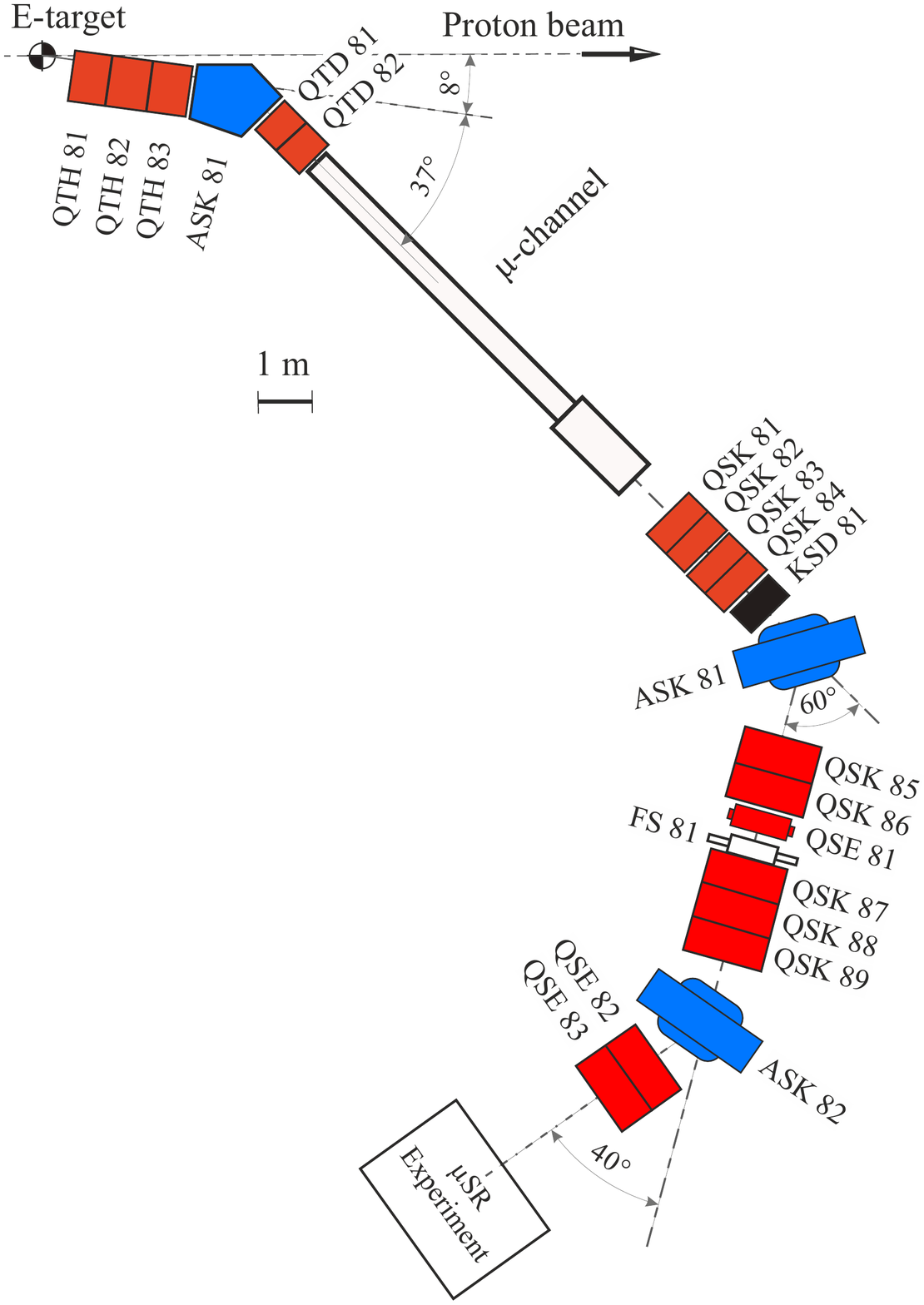}
\vspace{-1.5cm}
\caption{\footnotesize Schematic layout of $\mu$E1 beam line at PSI. The superconducting solenoid ($\mu-$channel) is the effective source of muons generated from pions decaying in flight inside it.  Quadrupolar magnets (Q) focus the beam and bending magnets (A) select the momentum. The slits (F) limit the lateral extension, the intensity and the momentum width of the muon beam.}
 \label{Fig:mue1}
\end{figure}

Fig. \ref{Fig:mue1} shows the $\mu$E1 decay line.
The 'beam optics' consist of magnetic elements to transport and focus the beam (quadrupoles, Q) to the instrument and select the momentum (dipoles A). Slits (F)  are used to adjust beam size, intensity and momentum width.
Pions produced in the thick target E in the forward direction are extracted  at 8$^{\circ}$ and collected over a small solid angle  using a triplet of half quadrupoles. The first dipole magnet ASK81 selects the momentum of the pions, which enter a 8~m long decay channel section consisting of a superconducting solenoid with a longitudinal field of $\simeq 5$~T.  A large fraction of pions with not too high momentum decay in the channel. In this respect the solenoid may be viewed as an extended muon source. The second and third dipole magnets together with the slit system perform a muon momentum and momentum width selection. Usually, only the muons emitted parallel (forward) or antiparallel (backward) to the muon momentum are selected, in order to obtain a large degree of polarization, which in the laboratory frame is limited to about 80\%. Note that opposite to the 'surface` muons having spins oriented antiparallel to their momentum, the 'backward` muons generally used for pressure $\mu$E1 have spins parallel to the momentum (see also Figs.~\ref{Fig:MSR_schematic} and \ref{Fig:TF-LF}).
The characteristics of $\mu$E1 beam line are summarized in Table~\ref{Table:mue1}.

\begin{table}[htb]
\caption[~]{\label{Table:mue1} Characteristics of $\mu$E1 beam line at PSI used for high-pressure experiments. }
\begin{center}
 \vspace{-0.5cm}
\begin{tabular}{llccccccccccc}\\
\toprule
Momentum acceptance (FWHM)& 3\% \\
Pion momentum range [MeV/c]& 200--125\\
Muon momentum range [MeV/c]& 125--60\\
Rate of positive muons [mA$^{-1}$s$^{-1}$]& $2\cdot 10^8-1\cdot10^8$\\
Rate of negative muons [mA$^{-1}$s$^{-1}$]& $6\cdot 10^7-3\cdot10^7$\\
Spot size (FWHM) & $39\times 25$~mm\\

\botrule
\end{tabular}
   \end{center}
\end{table}

\subsubsection{The General Purpose Decay spectrometer (GPD)}

\begin{figure}[t!]
\centering
\includegraphics[width=1.0\linewidth]{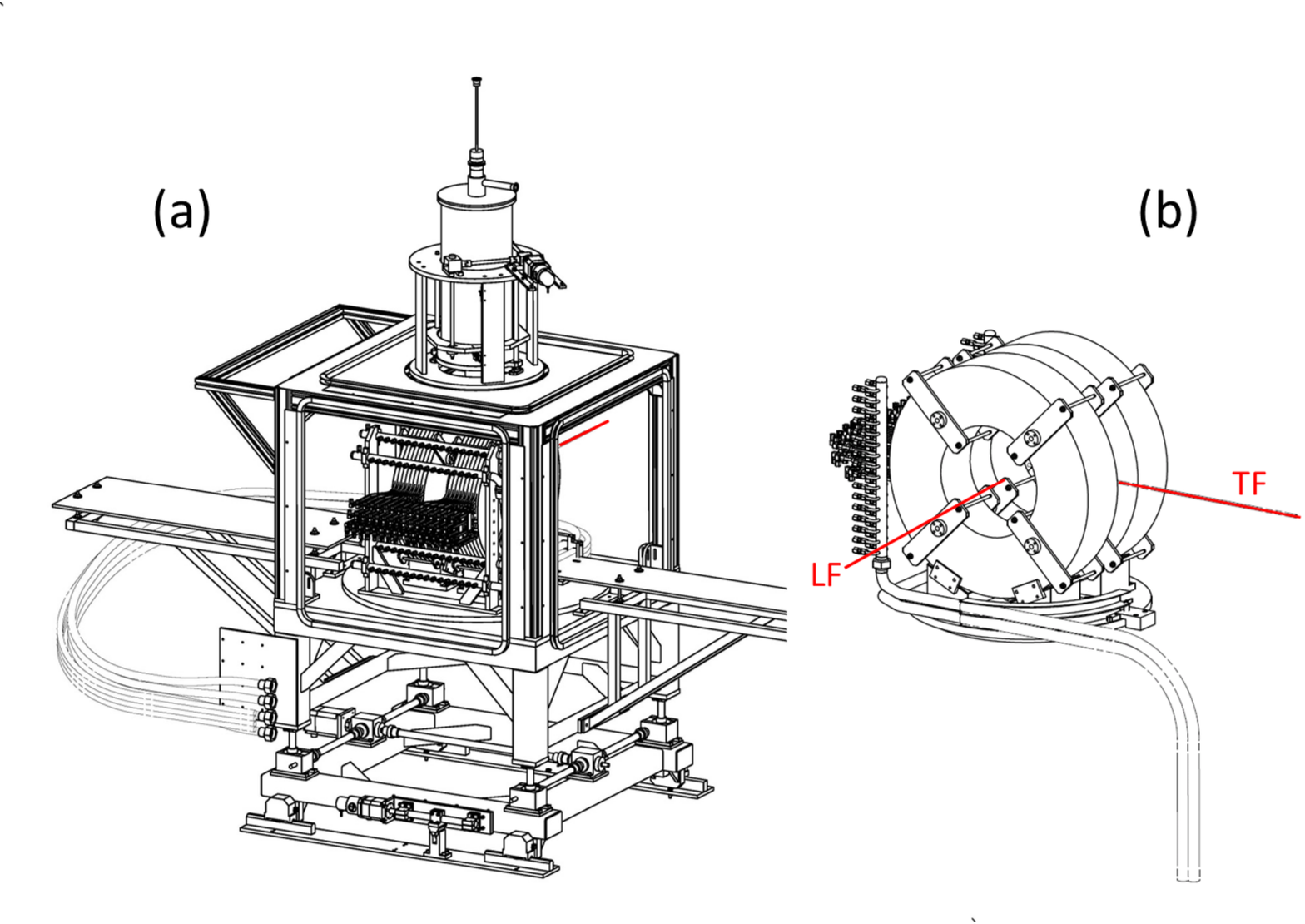}
\caption{\footnotesize (a) 3D view of the GPD instrument at PSI. (b) The Helmholz coils magnet: the red lines are the possible muon beam directions. In order to perform the transversal field (TF) and the longitudinal field (LF) experiments the magnet could be rotated by 90 degrees. }
 \label{Fig:GPD}
\end{figure}

The GPD $\mu$SR spectrometer (Fig.~\ref{Fig:GPD}~a) is permanently installed in the $\mu$E1 area. The instrument is designed to perform $\mu$SR experiments in zero-field (ZF), longitudinal-field (LF), and transverse-field (TF) in a wide temperature range making use of dedicated cryostats (see Table~\ref{Table:Cryostat}). The GPD spectrometer is equipped with a water cooled Helmoholz coils magnet providing a maximum field $\simeq 0.6$~T. To select between the longitudinal  and the transverse field geometry the magnet can be turned by 90 degrees (Fig.~\ref{Fig:GPD}~b).  Sample rotation is provided for angular dependent studies in single crystals. A plastic degrader can be installed at the entrance to reduce the energy of the incoming muons. The size of the incoming muon beam is defined by passive lead collimators. Cylindrical collimation with $\varnothing$ 16, 12, 10, 8, or 6~mm and rectangular collimation over $4\times 10$~mm$^2$ is possible.

\begin{table}[htb]
\caption[~]{\label{Table:Cryostat} Sample cryostats available at the GPD instrument. }
\begin{center}
 \vspace{-0.5cm}
\begin{tabular}{llccccccccccc}\\
\toprule
Cryostat& Brand& $T-$range& Use with \\
&&&pressure cells\\
\colrule
$^3$He Sorption pumped & Oxford&0.24-325~K& Yes \\
$^4$He gas flow& Janis& 2.5-300~K & Yes\\
Closed Cycle Refr.& Home made & 10-300~K & No\\
N$_2$ gas flow& Home made& 80-500~K& No\\

\botrule

\end{tabular}
   \end{center}
\end{table}

\begin{figure}[htb]
\centering
\includegraphics[width=0.8\linewidth, angle=0]{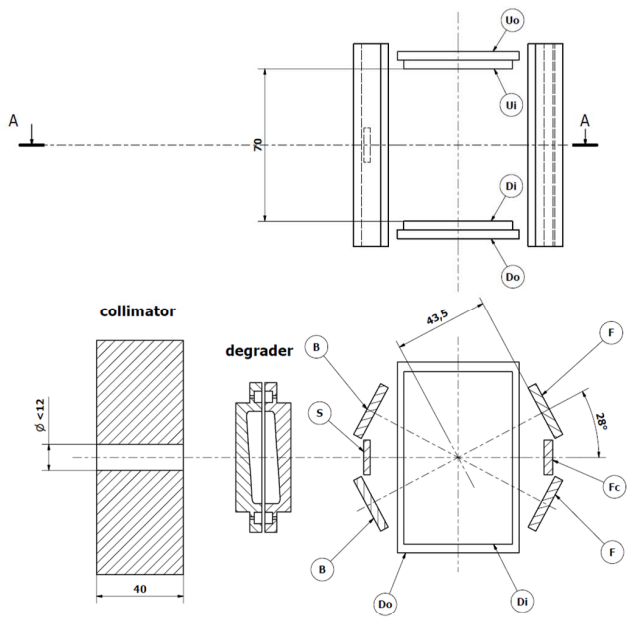}
\caption{\footnotesize The schematic view of the detectors at the GPD spectrometer. The meaning of the detectors are:  S -- start,  F -- Forward, FC -- Forward center, B-- Backward, U -- Up, and D --Down. The U and D detectors consist of two detectors each (i -- inner and o -- outer). The collimator reduces the size of the incoming muon beam. The degrader allows to decrease the momentum of the incoming muons. }
 \label{Fig:Detectors}
\end{figure}

The detectors consist presently of plastic scintillators. The detectors are arranged as (i) a muon counter (start detector S,
3.0~mm thick) and (ii) five positron detectors denominated (with respect to the beam direction) Forward (F), Forward center (FC), Backward (B), Up (U), and Down (D) (see Fig.~\ref{Fig:Detectors}). The U and D detectors are divided in two subdetectors (U$_{\rm o}$, U$_{\rm i}$ and D$_{\rm o}$, D$_{\rm i}$). Each of the detector is read by a photomultiplier tube (Hamamatsu-PTs R1828-01). The forward center detector can be used in the so-called ''veto`` mode for detecting muons (and their decay positrons) which have not stopped in the sample and rejecting the corresponding events. It is used in experiments without the pressure cell. With the pressure cell installed, when all the muons are stopped in the cell and the sample, the FC detector is combined with the F detector to increase the forward solid angle.

\subsection{$\mu$SR under pressure} \label{Sec:muSR_under_pressure}

The asymmetry function described by Eq.~\ref{eq_muSR_asymmetry} contains the information on the physics of the sample studied. Due to the local character of $\mu$SR the asymmetry has ''additive`` character, in the sense that muons probing different local magnetic environments will have a distinctive signature each contributing to the asymmetry spectrum:
\begin{equation}
A_0 P(t)=\sum_i A_i P_i(t), \ {\rm with} \ A_0=\sum_i A_i.
 \label{eq:Asymmetry_common}
\end{equation}
This property allows to determine volume fractions of different phases and their interaction.
For a $\mu$SR pressure experiment, the measured asymmetry contains the separate contributions of the pressure cell as a background signal and of the sample, which has to be maximized:
\begin{equation}
A_0 P(t)=A_{pc} P_{pc}(t)+A_{s} P_{s}(t), \ {\rm with} \ A_0=A_{pc}+A_s,
 \label{eq:Asymmetry_PC}
\end{equation}
(where '{\it pc}' and '{\it s}' denote the pressure cell and the sample, respectively).

\begin{figure}[htb]
\centering
\includegraphics[width=0.6\linewidth]{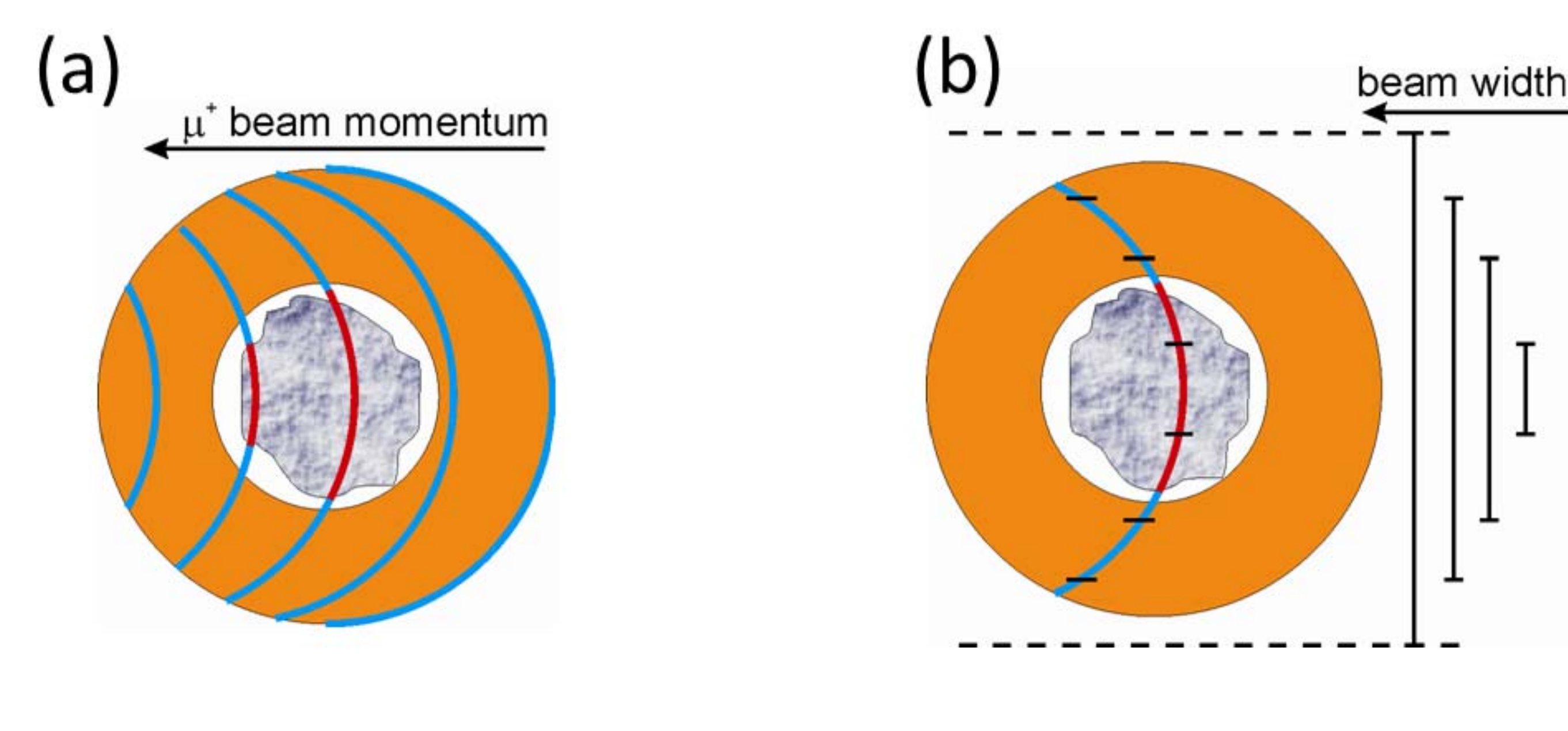}
\caption{\footnotesize  Transverse section of a piston-cylinder pressure cell used for $\mu$SR pressure experiments. The sample, grey object, is in the center. 
(a) Stopping profiles of the muons inside the sample (red) and the cell (blue) for different values of the $\mu^+$ momentum. The width of the beam is chosen to be larger than the diameter of the pressure cell. The optimum value of the momentum maximizes the ratio between the number of muons stopped inside the sample and the pressure cell walls. (b) Stopping of the muons inside the sample and the pressure cell for different values of the beam width (beam spot size).}
 \label{Fig:muons_in_PC}
\end{figure}

By tuning the muon momentum, one chooses the value maximizing the ratio between the number of muons stopped inside the sample and the number of the muons stopped in the pressure cell walls (signal to background ratio). As illustrated in Figs.~\ref{Fig:muons_in_PC}~(a) and (b) this ratio depends not only on the muon momentum  but also on the lateral extension of the muon beam, which ideally should be as small as possible to reduce the contribution of muons missing the sample while stopping in the cell walls.

However, the size of the muon beam cannot be arbitrarly reduced. The finite dispersion of the beam momentum has to be convoluted with the 'natural` longitudinal and transverse straggling of the muons as a consequence of 
energy loss and scattering while crossing the shields of the cryostat, pressure cell walls and thermalizing in the sample. These effects amount to a spread of the order of a few millimeters, and limit the minimum possible dimension of the beam spot at the sample.
Moreover, the size of the beam is also momentum dependent. Obtaining the optimum values of the beam momentum, which also depends of the particular choice of the pressure cell (see below) is therefore a key operation, which is performed at the beginning of a measuring series.
Summarizing:
\begin{itemize}
  \item Samples of relative large dimensions are required ($\sim$ 150 mm$^3$) in order to have a sizable signal from the sample inside the pressure cell. This is contrast to other techniques such as resistivity, AC and DC magnetization, specific heat, {\it etc.} where samples of less than one mm$^3$ can be measured thus allowing the use of {\it e.g.} anvil type of cells.

  \item At present, the only practical choice for $\mu$SR experiments are piston-cylinder pressure cells, which provide a reasonable amount of sample space \cite{Klotz_book_2013}. Development at PSI has concentrated on improving the characteristics of this type of cells within the boundary condition to use them in an existing sample and cryogenic environment (which limits the maximum outer diameter of the pressure cell to about 26~mm).

  \item Our experimental tests reveal that more than 50\% of the muons can be stopped in the sample by using a piston-cylinder pressure cell with $\varnothing 7$~mm inner diameter and about 30\% for $\varnothing 5$~mm cell. On the other hand, $\varnothing 7$~mm cells sustain less pressure than the $\varnothing 5$~mm ones. Thus the choice of the specific pressure cell will be dictated by the experimental requirements.

\end{itemize}

\section{Pressure cells for $\mu$SR experiments} \label{sec:Pressure_cell}

Several parameters are determinant to perform successful $\mu$SR measurements under pressure. The most important are the availability of a large sample volume and hydrostatic pressure conditions. In addition, low temperatures ({\it i.e.} sub-Kelvin) are generally required to study several relevant condensed matter topics. Another factor is the possibility to perform pressure scans without actually opening the cell. In addition to the obvious time saving, which is essential when using a costly particle beam, this latter point is important (i) for calibration reasons: each opening of the cell may lead to a slight change of the sample position, and/or cause a redistribution of the sample pieces in case of a multi-parts sample, which would require a new time-consuming calibration; (ii) the physical properties of the sample may change during each fresh pressurization.
All the above mentioned points require the use of piston-cylinder pressure cells for which the mechanical properties can be either calculated using the long cylinder theory \cite{Klotz_book_2013, Eremets_book_1996} and/or simulated by using finite element analysis applications (see {\it e.g.} Ref.~\cite{Wang_RSI_2011} describing pressure cells used for inelastic neutron scattering and Fig.~\ref{Fig:ANSYS} for an example of the PSI
double-wall pressure cell open on both sides).

In the following Sections, we will briefly overview the history of $\mu$SR experiments performed under pressure, describe the design of the pressure cells used at PSI and present the properties of the materials used to build the pressure cell and related accessories.

\subsection{$\mu$SR experiments under pressure -- brief historical overview}

The first $\mu$SR experiments under pressure at PSI were performed in 1980 by Butz {\it et al.} \cite{Butz_PLA_1980}. A clamped cell with oil as pressure transmitting media (maximum pressure, $p_{max}\simeq 0.7$~GPa) was used for studying the transition metals Fe and Ni.
As a next step, in 1986 a gas pressure cell was designed ($p_{max}\simeq 1.4$~GPa, Helium as pressure transmitting media) \cite{Butz_HI_1986, Kratzer_HI_1994}. However, due to safety reasons, the maximum pressure available for these cells was limited to about 0.9~GPa. This type of pressure device was successfully used until 1999 for various studies  \cite{Butz_HI_1986, Kratzer_HI_1994, Hartmann_HI_1994, Wackelgard_HI_1986, Schreier_HI_1997, Schreier_PHD_1999, Martin_PHD_1996, Martin_PhysB_2000, Kalvius_HI_2000}. In addition to the limited pressure range, these cells had the disadvantage that the Helium capillary imposed quite severe temperature range restrictions, with typical lowest temperatures slightly below 10~K. Moreover, safety reasons imposed a release of the pressure when accessing the experimental area which caused unwanted pressure cycles.

At the beginning of 2001, a new generation of PSI pressure cells came into operation \cite{Andreica_thesis_2001}. The starting point was a close collaboration with the group of D.~Jaccard from the University of Geneva. A first development was a single layer piston cylinder cell, open solely at one end, using liquid as a pressure transmitting media and made of a Copper Beryllium (CuBe) alloy.
The peculiar opening at only one end was adopted because of the restrictions imposed by the sample environment, {\it i.e.} the distance between the center of the cryostat windows to the bottom of the sample space.
This cell could reach  $\simeq 0.9$~GPa ($\simeq1.4$~GPa) for a 7~mm (5~mm) inner diameter of the cell.

A further step was accomplished with the use of MP35N (Ni-Co-Cr-Mo) alloy for the body of the pressure cell and pressures up to about 1.9-2.0~GPa could be routinely obtained.
The latest developments are double layer cells made of a combination between CuBe and MP35N. To guarantee a $\mu$SR signal with a high enough signal to background ratio, the dimensions of the sample space are rather large limiting therefore the maximum pressure to 2.5~GPa (1.8~GPa) for a sample space with a diameter of 6 mm for pressure cells  opened from both sides made out of MP35N/MP35N (or CuBe/CuBe, respectively). This type of cells will be presented in more details in Section~\ref{Sec:P-Cell}.

It is fair to say that presently the bulk of $\mu$SR experiments under pressure, as well as of the further development of $\mu$SR pressure cells are performed at PSI. This is due to the unique combination of high-pressures and sub-Kelvin croygenics capabilities available at the GPD instrument. Nevertheless, some activity takes place also at ISIS (UK) and at TRIUMF (Canada). At ISIS the experiments are carried out using the high-momentum RIKEN-RAL beamline by using gas pressure cell with $p_{max}\simeq 0.6$~GPa \cite{Watanabe_PhysB_2009, Telling_PRB_2012, Ellis_PPr_2012, Enomoto_PolHedr_2011, Suzuki_JPCS_2010a, Suzuki_JPCS_2010b, Suzuki_PRB_2009}.
At TRIUMF a single wall piston-cylnder type of pressure cell with maximum pressure $\simeq 2.3$~GPa is used. The experiments are conducted  at the M9B beam-line by using the dedicated `Omni-Prime' spectrometer.

\subsection{Double-wall cells used at PSI} \label{Sec:P-Cell}

\begin{figure}[b]
\centering
\includegraphics[width=0.7\linewidth, angle=0]{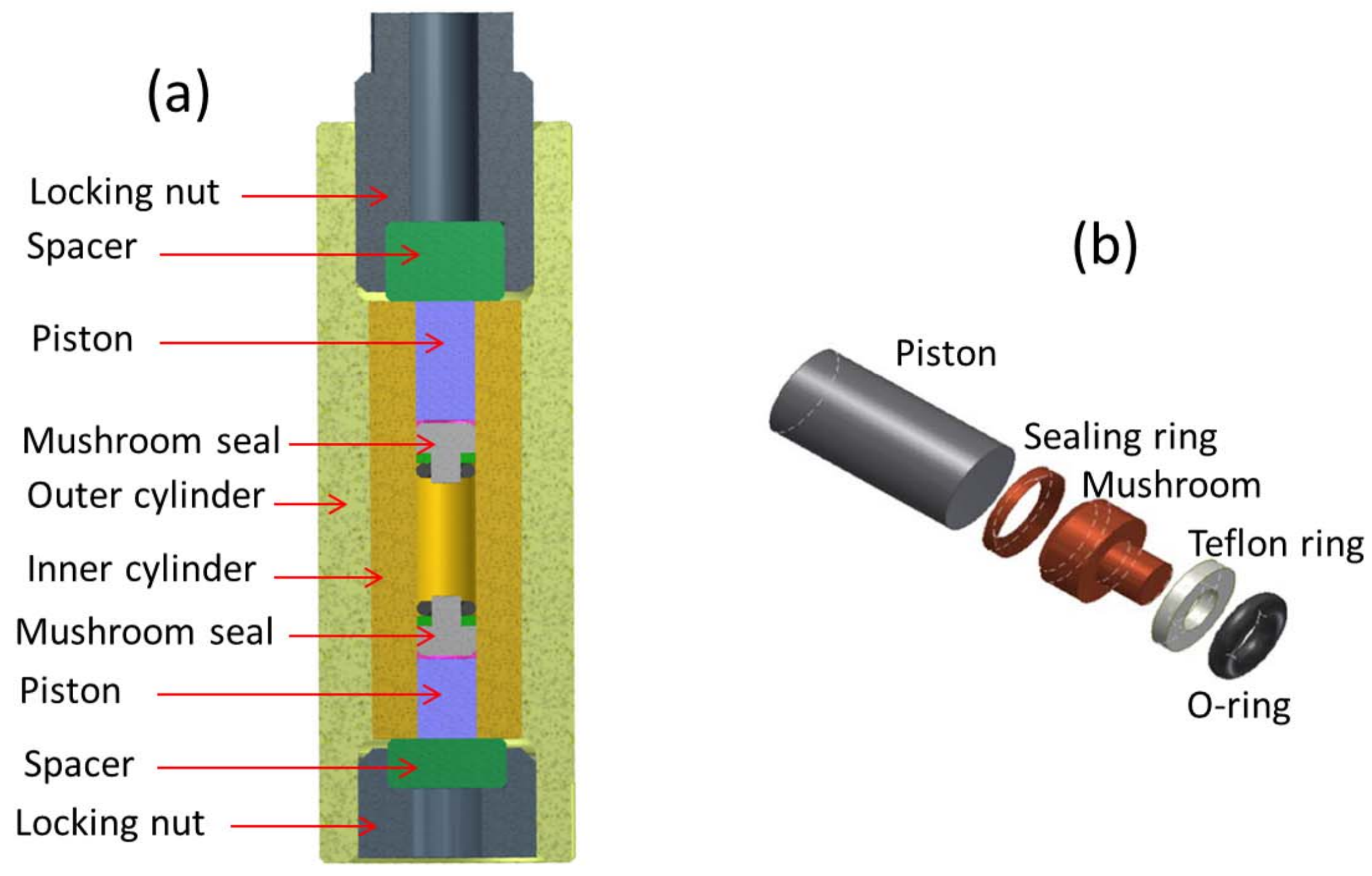}
\caption{\footnotesize (a) Cross-sectional view of the PSI pressure cell; (b) Details of the pressure seal. }
 \label{Fig:p-cell}
\end{figure}

The cross-sectional view of a fully assembled typical double-wall pressure cell is presented in Fig.~\ref{Fig:p-cell}. The main components of the cell are: (i) a cylindrical double-wall body, (ii) pistons, (iii)'mushroom` seals, (iv) locking nuts, and (v) spacers. The latest are  needed to support the pistons and prevent their rotation when the nuts are tightened. Diameters and heights of the outer and the inner cylinders are $\varnothing=24$~mm, $h=72$~mm, and $\varnothing=15$~mm, $h=44$~mm, respectively. Two types of cells with inner diameter 6~mm and 7~mm were produced. With both pistons completely inserted, the maximum sample height is $\simeq 12$~mm.
The body of the cell, the top and the bottom locking nuts and the mushroom pieces are made of MP35N alloy (Ni -- 35\%, Co -- 35\%, Cr--20\%, and Mo--10\% in weight percent, see Ref.~\cite{Carpenter}). All the pieces were heat-treated for 4 hours at 590~$^{\rm o}$C after machining. Pistons and spacers are made of non-magnetic tungsten carbide ({\sc ceratizit-ctf21r}, see Ref.~\cite{Ceratizit}). The sealing rings were made of fully hardened Copper Beryllium alloy ({\sc berilco-25}).

The body of the pressure cell consists of two parts -- the inner and the outer cylinders which are shrink fitted into each other. Two alternative methods for the fabrication of the pressure cell  were tried. In the first approach, the inner cylinder was prepared with a diameter $\simeq 0.10$~mm bigger than the inner diameter of the outer cylinder. By heating the outer cylinder up to $590^{\rm o}$C and cooling down the inner one to liquid nitrogen temperature it is possible to fit them into each other. By coming back to ambient temperatures the inner cylinder remains under radial compression from the outer cylinder. In the other approach, the outer surface of the inner and the inner surface of the outer cylinders were machined conically with the angle of $\simeq 1^{\rm o}$. The diameter difference of both fully assembled cylinders is $\simeq 0.12$~mm. The pressure cell is produced by mechanically inserting the inner cylinder into the outer one by using a hydraulic press. This process is similar to the situation of a cold/warm fitting of the cylinders as described above. The friction between the two parts is high enough to keep the cell fully assembled.
Note that pressure cells produced by both described ways were found to behave very similarly with respect to the the maximum reachable pressure, as well as with respect to the pressure loss caused by friction between the sealing system and the pressure cell walls.

\begin{figure}[htb]
\centering
\includegraphics[width=1.0\linewidth, angle=0]{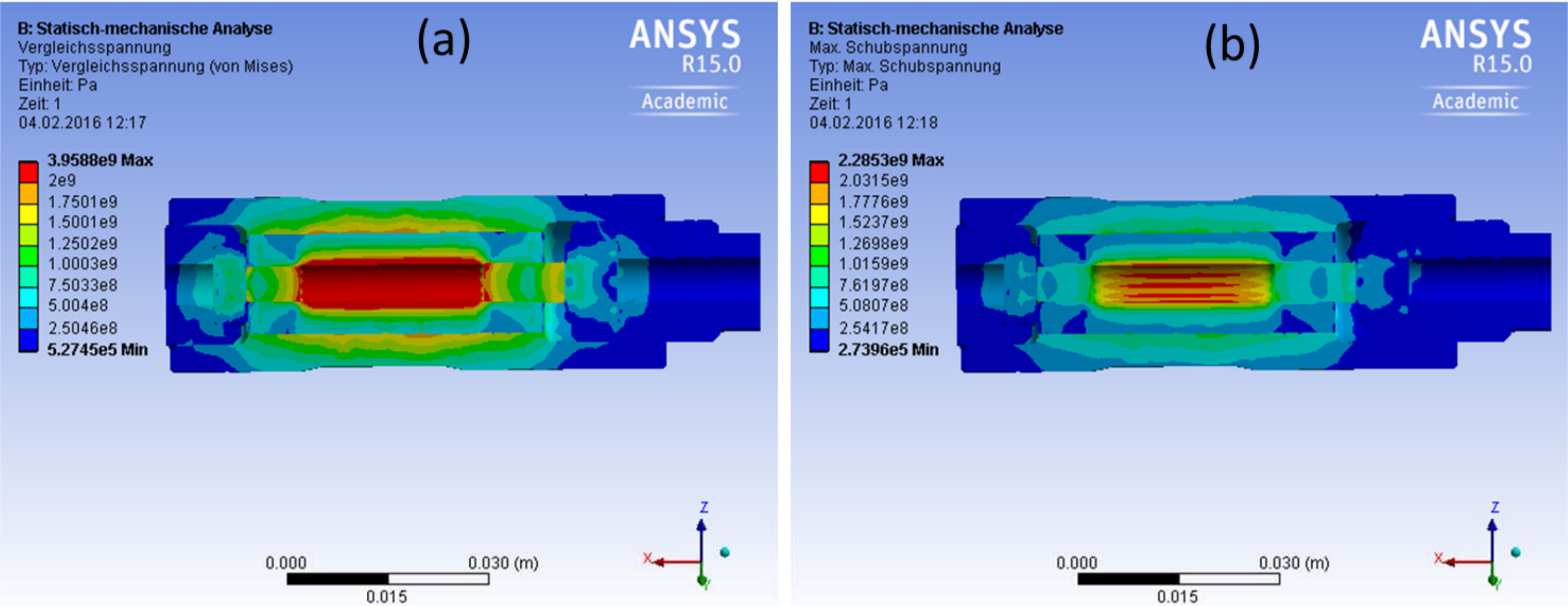}
\caption{\footnotesize Maximum principal stress  (panel a) and shear stress (panel b) in a double-wall MP35N pressure cell used in $\mu$SR experiments. The calculations are made using the  {\sc ansys r15.0} software. }
 \label{Fig:ANSYS}
\end{figure}

In order to investigate the mechanical response of the cell to the applied pressure, a finite element analysis by using the {\sc ansys r15.0}  software was performed \cite{ANSYS}
using the appropriate boundary conditions and including a modeling of the force between two cylinders.
Figure~\ref{Fig:ANSYS} shows a sectional view of the body of the cell with the corresponding principal stress and the shear stress distributions. These calculations were made for `extreme' conditions, {\it i.e.} when the stress at the border between two cylinders reaches the so-called ultimate tensile strength of the MP35N alloy (2.07~GPa, Ref.~\cite{Carpenter}). The results reveal that the `theoretical' limit for a double-wall cell made of MP35N alloy is of the order of 3.9~GPa.
For safety reasons, the maximum applied pressure is always limited to 3.5~GPa. The applied pressure at closing time is calculated as $p_{\,\text{closing}}=F_{\,\text{Load}}/S$, where $F_{\,\text{Load}}$ is the force applied by the hydraulic press and $S$ is the surface area of the piston. Note that $p_{\,\text{closing}}$ is usually much larger than the effective real pressure inside the cell due to friction effects.

As will be shown later for $T\lesssim 1$~K, the $\mu$SR response of a MP35N cell is strongly temperature dependent. Therefore, at very low temperature, the cell signal may be difficult to separate from the one of the sample. For precise experiments we have also prepared a single wall pressure cells of a similar design (as in Fig.~\ref{Fig:p-cell}) made of Copper Beryllium alloy. Even though such a cell allows to reach a maximum pressure just half of the one accessible with a double wall MP35N cell, the paramount advantage of it is a $\mu$SR response that is temperature independent at very low temperatures.

\subsection{Pressure seal design}

The design of the seal is defined by two requirements. First, as shown in Section~\ref{Sec:muSR_under_pressure}, in $\mu$SR experiments under pressure the muons stop not only in the sample but also in the pressure cell walls. Therefore, the muon asymmetry spectra described by Eq.~\ref{eq:Asymmetry_PC} consists of `sample' and `pressure cell' contributions and one should avoid the presence of any other material, except the sample, inside the pressure cell channel. A second requirement is to have the possibility to perform several pressure changes without opening the pressure cell.

We approached the problem by implementing the seal shown in Fig.~\ref{Fig:p-cell}b.  The seal assembly consists of a conically shaped sealing ring (hardened Copper Beryllium), a mushroom-shaped piece (MP35N alloy), a teflon ring and a rubber O-ring.
The initial sealing is made by the O-ring allowing one to close the cell and to increase the pressure up to the level where the teflon ring starts to flow and to fill the gap between the mushroom and the pressure cell walls.
The conical Copper Beryllium ring impedes that the teflon flows further which would cause a blocking of the piston. This design makes full use of the bore of the cell to accommodate the sample, and allows to perform repeated pressure changes  in increasing or decreasing directions.

\subsection{Pressure cell loading, pressure determination and pressure loading curve}

The process of loading the cell is permanently monitored by using a specially designed computer-controlled system allowing to measure in-situ the displacement of the piston ($x$) and the radial expansion of the pressure cell as a function of loading force $F_{\,\text{Load}}$. Special attention is paid to maintain the pressure cell in the `elastic' regime, {\it i.e.} by ensuring that the slope ${\rm d}x/{\rm d} F_{\,\text{Load}}$ of the loading curve $x( F_{\,\text{Load}} )$ is almost constant. A sudden change of the slope ${\rm d}x/{\rm d} F_{\,\text{Load}}$ would correspond either to a leakage of the cell or would signal that one exceeds the local stress level above the elastic regime.
In both cases the loading process must be immediately stopped and the pressure in the cell released  to zero.

\begin{figure}[t]
\centering
\includegraphics[width=0.8\linewidth, angle=0]{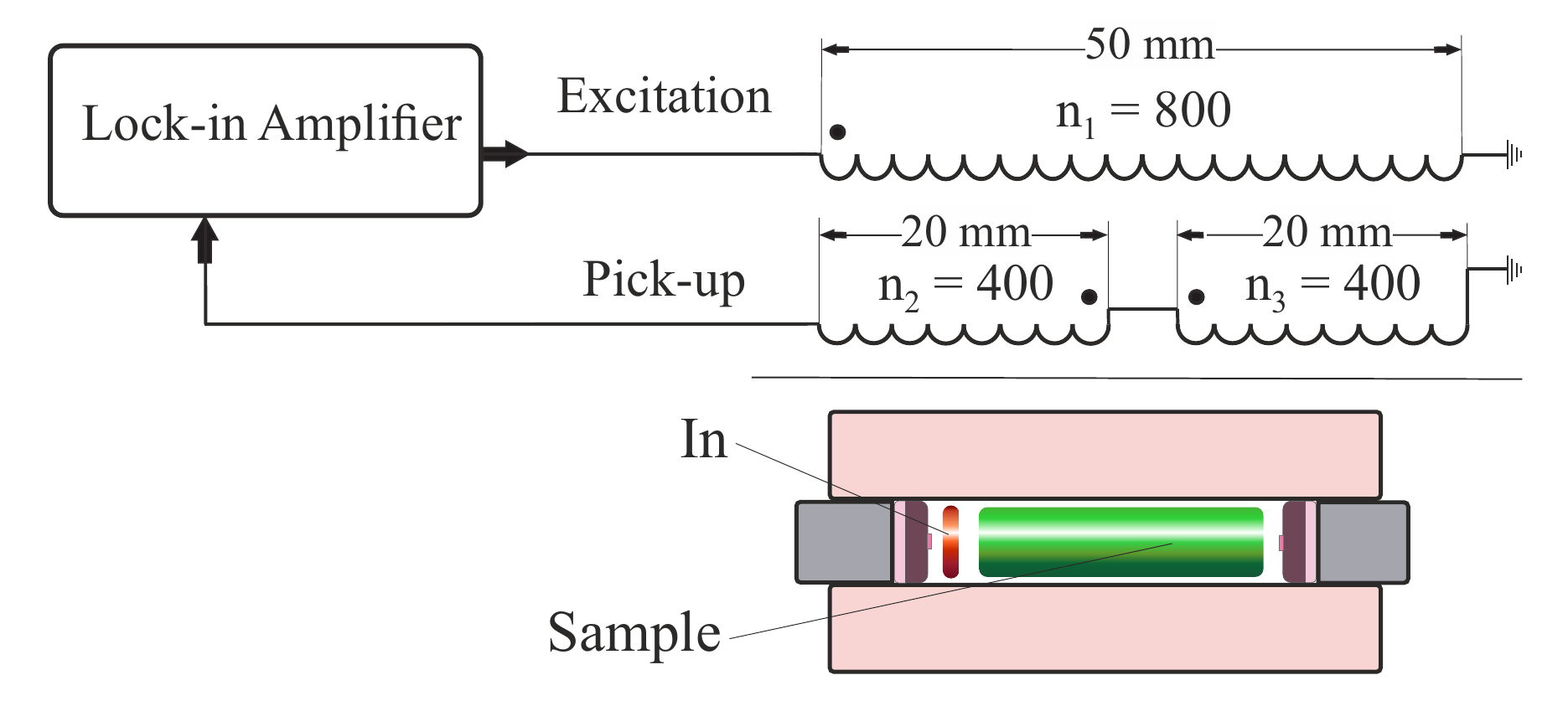}
\caption{\footnotesize Schematic diagram of the pressure measurement system. The AC coils consisting of the excitation and two pickup coils is mounted on a separate cylindrical holder with inner diameter 27~mm. The cell is placed inside the coil in way that the In pressure indicator is located at the center of one pick up coil. The transition of In into the superconducting state leads to the abrupt change output AC voltage. }
 \label{Fig:Lockin}
\end{figure}

The exact pressure inside the cell is determined by monitoring the pressure-dependent shift of the superconducting transition temperature ($T_c$) of a small piece of Indium (pressure probe).
The volume of the pressure probe is much lower than the one of the sample to avoid an additional background contribution to $\mu$SR signal. For the superconducting transition temperature of Indium the following relation holds \cite{Eiling_JPF_1981}:
\begin{equation}
T_c(p) = T_c(0) - 0.3812 \; p + 0.0122\;  p^2,
 \label{eq:Tc-In}
\end{equation}
where $T_c(0)\simeq 3.40$~K is the transition temperature of In at ambient pressure.

The superconducting transition of the pressure probe is determined by AC susceptibility measurements. A coil surrounding the pressure cell (input coil) is fed with an AC current. A second coil, inside the first one, made up of two sub-coils (one centered at the position of the pressure probe) wound in opposition is used as pick-up coil. The signal of the pick-up coil is compensated above $T_c$ but becomes uncompensated when the superconducting transition is crossed upon lowering the temperature. The lack of compensation is detected using a lock-in amplifier which compares the phase and amplitudes of the input signal with the ones of the pick-up signal (see Fig.~\ref{Fig:Lockin}).
Note, that the pressure is determined at temperatures corresponding to $T_c$ of Indium ({\it i.e.} $\simeq 3$~K). By performing experiments at higher temperatures one needs to consider the temperature dependent pressure drop of the pressure transmitting medium (see {\it e.g.} Ref.~\cite{Torikachvili_arxiv_2015} and references therein).

\begin{figure}[t]
\centering
\includegraphics[width=0.6\linewidth, angle=-0]{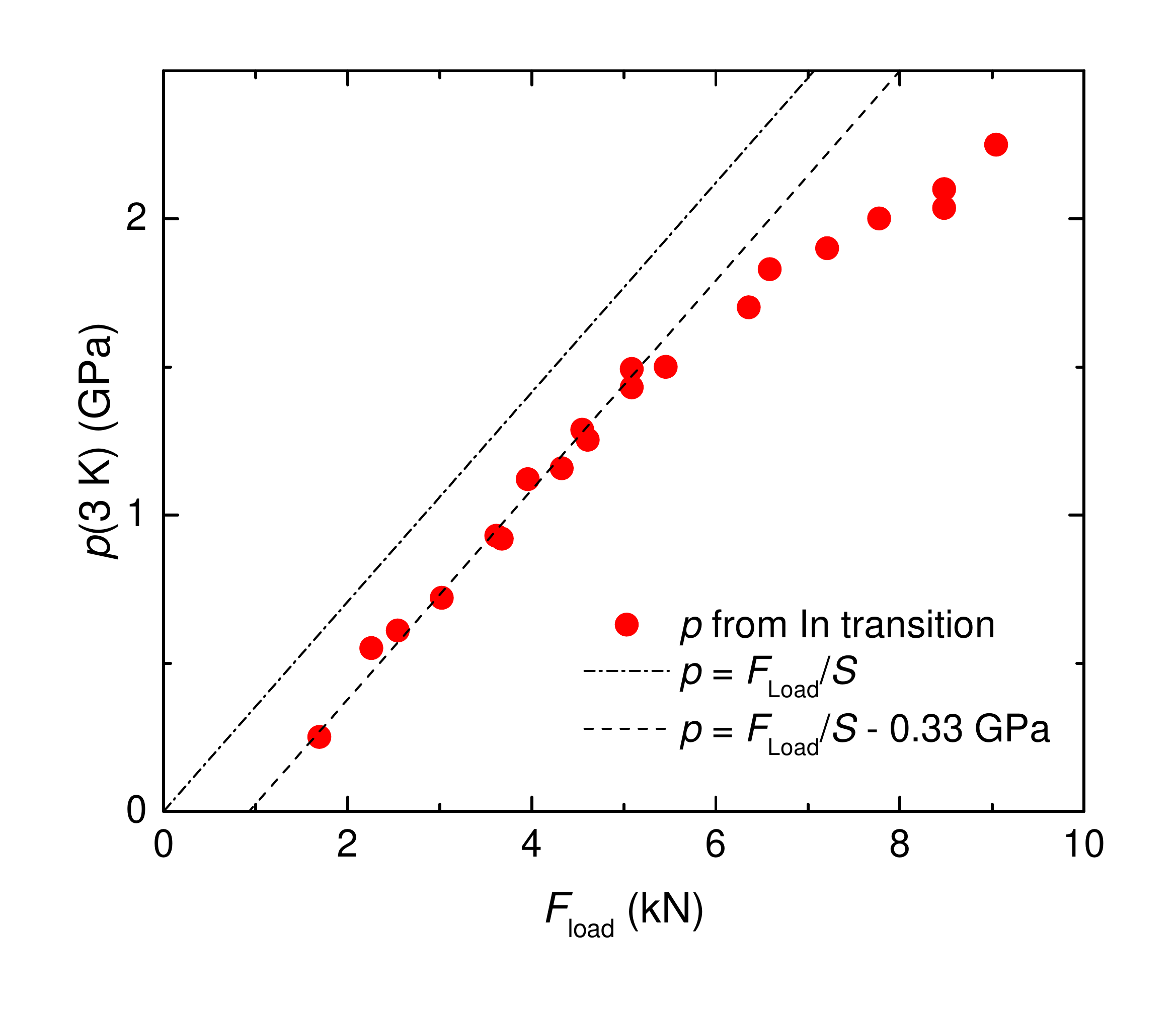}
\caption{\footnotesize Pressure-load performance of the cell. Filled circles are experimental
data. The pressure inside the cell is determined by for the pressure induced shift of the superconducting transition of Indium. The dash-dotted line is the ideal pressure calculated from the applied force ($F_{\,\text{Load}}$) over the area of the piston ($S$). The dashed line is the curve shifted by 0.33~GPa. The shift 0.33~GPa corresponds to the pressure drop caused by the temperature induced compression of the pressure transmitting medium (Daphne 7373 oil).  }
 \label{Fig:Pressure_drop}
\end{figure}

The measured pressure at low temperatures inside the cell is shown as a function of the load $F_{\,\text{Load}}$ in Fig.~\ref{Fig:Pressure_drop}. The dash-dotted line is the ideal pressure $p_{\,\text{closing}}=F_{\,\text{Load}}/S$. Up to the pressure of $\sim 1.5$~GPa the difference between the ideal curve and the measured data is constant ($\simeq 0.33$~GPa) and it is entirely determined by temperature compression of the pressure transmitting medium (Daphne 7373 oil in this case). At higher loads frictional effects become noticeable. At a pressure of $\sim 2.3$~ GPa the difference between the ideal and the determined pressure is $\simeq 0.9$~GPa. Part of it (0.33~GPa) is due to the temperature induced compression of the  pressure transmitting medium, while $\simeq 0.6$~GPa is caused by friction.

\subsection{$\mu$SR background signal  caused by the pressure cell body}

The Copper Beryllium and MP35N alloys were used to build the body of the pressure cell, both because of their strength and their well-defined background contributions to the $\mu$SR signal.

\subsubsection{Copper Beryllium}

\begin{figure}[t]
\centering
\includegraphics[width=0.6\linewidth, angle=-0]{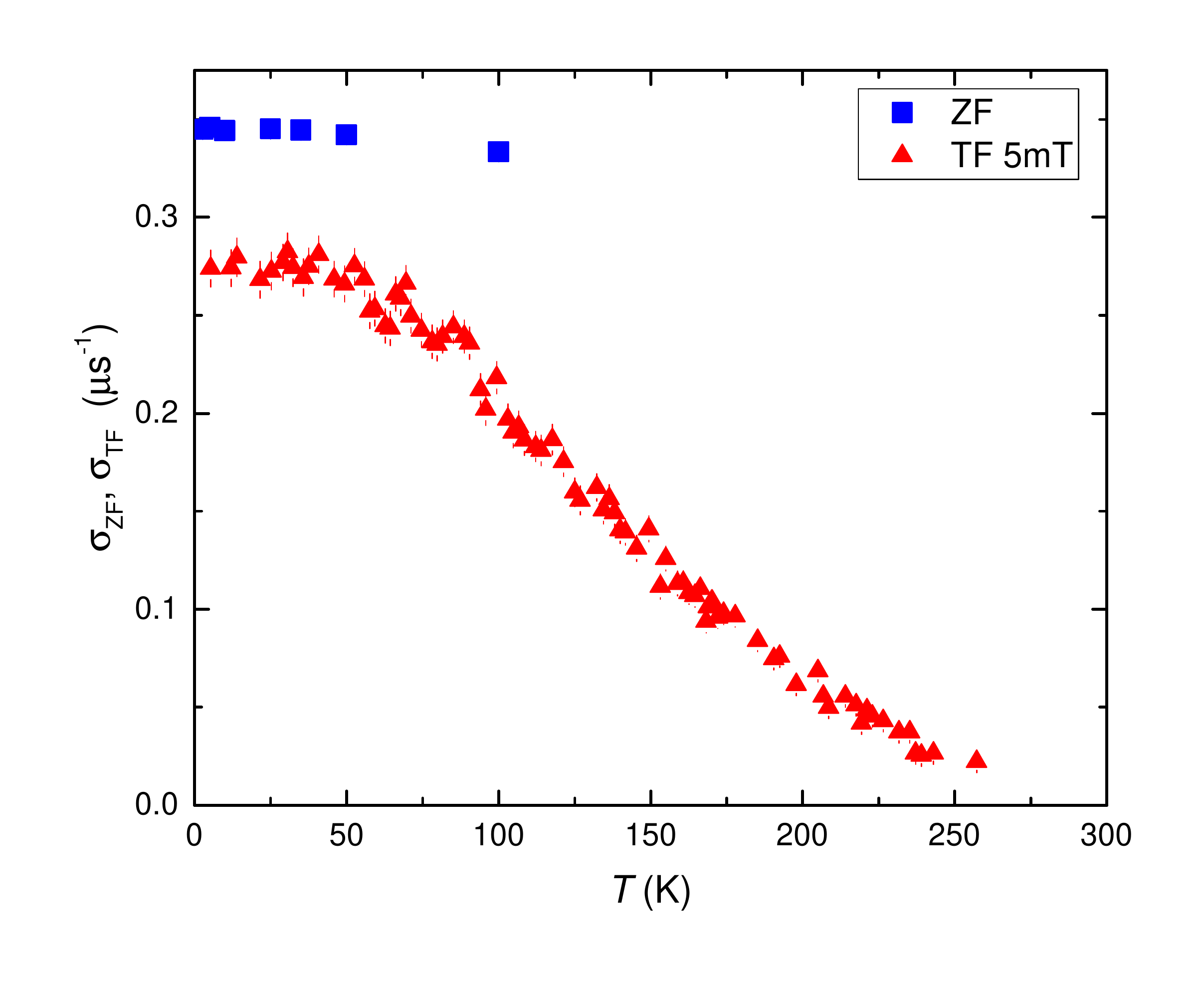}
\caption{\footnotesize Gaussian relaxation rate of the Copper Beryllium pressure cell measured in transverse (triangles) and zero-field (squares) modes.}
 \label{Fig:CuBe}
\end{figure}

Copper Beryllium ({\sc berilco-25}) is a nonmagnetic alloy down to at least 250~mK. Muons implanted in Copper Beryllium sense solely the magnetic field distribution created by the nuclear magnetic moments which can be considered static within the $\mu$SR time-window. Within these conditions, the zero-field (ZF) $\mu$SR spectra obtained on an empty cell made of Copper Beryllium can be fitted with a so-called Kubo-Toyabe depolarization function \cite{KuboToyabe_1967}:
\begin{equation}
P(t)_{\rm ZF}=\frac{1}{3} + \frac{2}{3} (1-\sigma_{\rm ZF}^2 t^2) \; e^{  -\sigma_{\rm ZF}^2 t^2/2 },
 \label{Eq:Gaussian-KT}
\end{equation}
where $\sigma_{\rm ZF}^2/\gamma_\mu^2$ represents the second moment of the field distribution along one cartesian axis perpendicular to the initial muon polarization. The temperature dependence of $\sigma_{\rm ZF}$ is reported  in Fig.~\ref{Fig:CuBe}.

The contribution of the cell in transverse-field (TF) experiments, where the external field $B_{ext}$ is applied perpendicular to the initial muon-spin polarization, is well described by a Gaussian depolarization function:
\begin{equation}
P(t)_{\rm TF}= e^{-\sigma_{\rm TF}^2 t^2/2} \cos(\gamma_\mu B_{ext} t + \phi)~.
 \label{Eq:Gaussian-TF}
\end{equation}
Here $\phi$ is a phase shift determined by the initial muon-spin  polarization with respect to the detectors geometry.
The second moment of the field distribution along the direction of the external field is given by $\sigma_{\rm TF}^2/\gamma_{\mu}^2$.
The temperature dependence of $\sigma_{\rm TF}$ measured with an applied field $B_{ext}=5$~mT is presented in Fig.~\ref{Fig:CuBe}.

\subsubsection{MP35N alloy}

\begin{figure}[htb]
\centering
\includegraphics[width=1.0\linewidth, angle=-0]{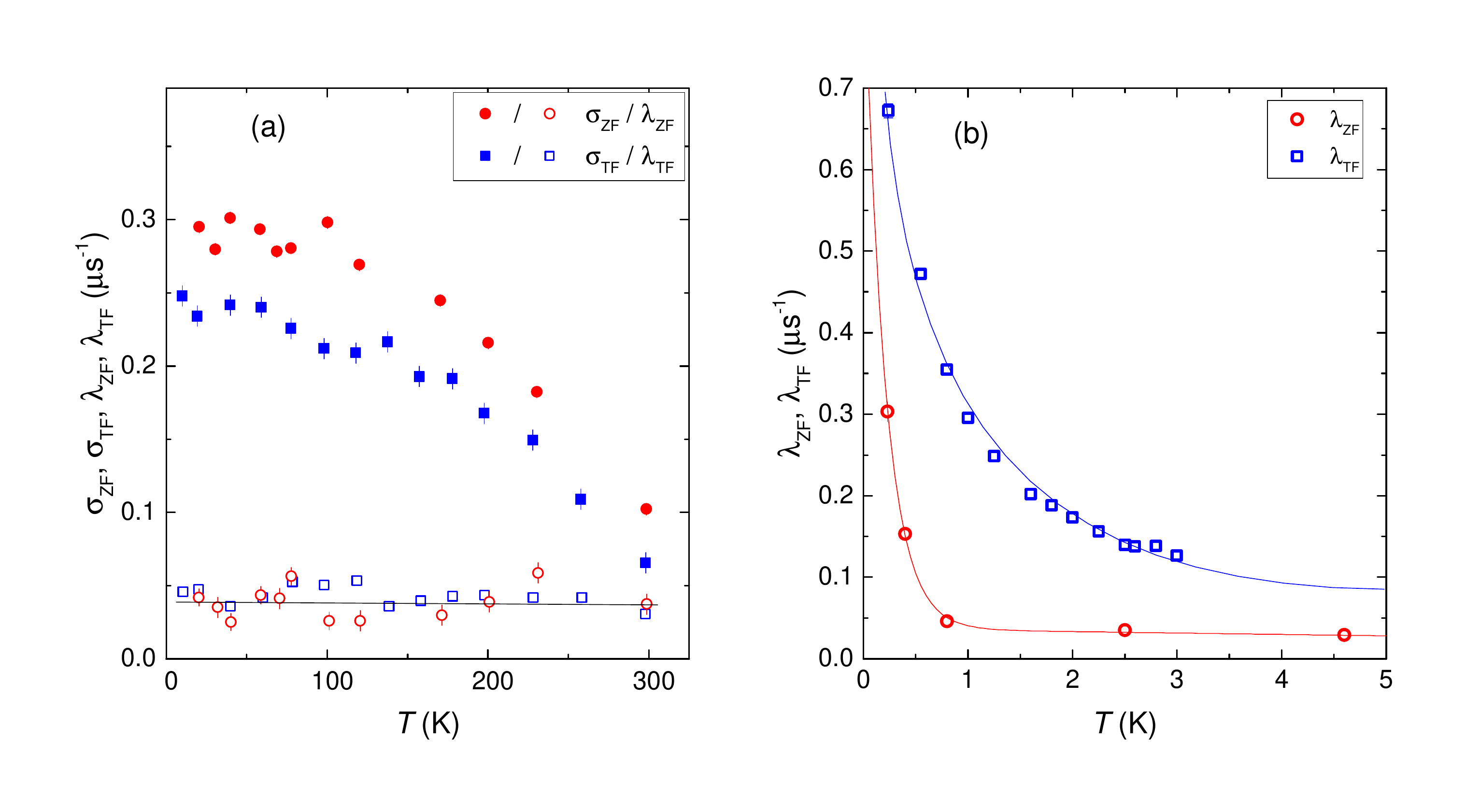}
\caption{\footnotesize (a) Gaussian ($\sigma$, full symbols) and exponential ($\lambda$, empty symbols) relaxation rates of MP35N measured in transverse- (applied field $B_{ext}=30$~mT, circles) and zero-field (squares) modes for $T$ above 5~K. The solid horizontal line
represents the mean value of $\simeq 0.04$~$\mu$s$^{-1}$. The data are adapted from Ref.~\cite{Maisuradze_PRB_2013}. (b) Temperature dependence of $\lambda_{\rm ZF}$ and $\lambda_{\rm TF}$ ($B_{ext}=50$~mT) measured at low temperatures. Note that $\lambda_{\rm TF}$ is field dependent. Lines are guides for the eye.}
 \label{Fig:MP35N}
\end{figure}

The alloy MP35N is nonmagnetic down to low temperatures, but its $\mu$SR response depends strongly on the applied field (see Ref.~\cite{Walker_RSI_99}). Muons implanted in MP35N sense both the static magnetic field distribution created by the nuclear magnetic moments and a dynamical field distribution created by the electronic moments in the paramagnetic phase. The ZF $\mu$SR spectra measured in the  MP35N alloy is therefore fitted by a Kubo-Toyabe depolarization function multiplied by an exponential damping:
\begin{equation}
P(t)_{\rm ZF}=\left[ \frac{1}{3} + \frac{2}{3} (1-\sigma_{\rm ZF}^2 t^2) \; e^{-\sigma_{\rm ZF}^2 t^2/2} \right] e^{-\lambda_{\rm ZF} t}.
 \label{Eq:Gaussian-KT-damped}
\end{equation}
Here $\sigma_{\rm ZF}$ and $\lambda_{\rm ZF}$ are the static (nuclear) and the dynamic (electronic)  relaxations rate, respectively. Results of ZF test experiments are presented in Fig.~\ref{Fig:MP35N}~(a). The depolarization rate $\lambda_{\rm ZF}$ is small down to 1~K then increases abruptly upon decreasing the temperature, signaling a slowing down of the fluctuations and indicating that correlations between the electronic moments start to build-up however without magnetic ordering down to 0.26~K, see Fig.~\ref{Fig:MP35N}~(b). This behavior may set some limitations for the use of MP35N pressure cells below 1~K.

In TF experiments the pressure cell contribution is analyzed by using the following functional form:
\begin{equation}
P(t)_{\rm TF}= e^{-\sigma_{\rm TF}^2 t^2/2} \; e^{-\lambda_{\rm TF} t} \cos(\gamma_\mu B_{ext} t +\phi).
 \label{Eq:Gaussian-damped-TF_}
\end{equation}
Temperature dependences of $\sigma_{\rm TF}$ and $\lambda_{\rm TF}$ are presented in Fig.~\ref{Fig:MP35N}.
Experience showed  that different parts of the MP35N rods, from which the pressure cell parts are produced, have slightly different chemical compositions and/or homogeneity, leading to slightly different depolarization rates. Hence, accurate $\mu$SR characterization measurements need to be performed separately for each particular pressure cell made of MP35N alloy.

\subsection{Pressure transmitting medium}

Hydrostaticity at low temperature is the key feature for a liquid used as a pressure transmitting medium (PTM). The reason is that a hydrostatic pressure is a thermodynamic parameter and the results obtained under such conditions 
 can readily be compared with theory. The most commonly used PTM's are either mixtures of 1:1 isopentane/n-pentane; 4:6 light oil/n-pentane; 1:1 and 5:1 isoamyl alcohol/n-pentane;  4:1 ethanol/methanol; 1:1 Fluorinerts FC72/FC84, FC84/FC87 and FC75/FC77;  or  `single' phase liquids like Daphne (7373, 7474 and 7575) oils, various Silicone oils and isopentane.
Among them, the mixtures of isoamyl alcohol/n-pentane and ethanol/methanol as well as the Daphne oils 7474 or 7575 remain hydrostatic at room temperature up to pressures $\gtrsim 4$~GPa, which is about the physical limit for the highest reachable pressure using  piston-cylinder cells \cite{Klotz_book_2013,Uwatoko_JPCM_2002}. The other PTM's are solidifying at lower pressures.
Comprehensive studies of these and other PTM's, with a special attention payed on the effects of non-hydrostaticity at high and low temperatures, are presented in Refs.~\cite{Klotz_book_2013, Klotz_JPD_2009, Tateiwa_RSI_2009, Torikachvili_arxiv_2015}.

For our $\mu$SR experiments under pressure either  Daphne 7373 oil or a 1:1 mixture of n-pentane/isoamyl alcohol are commonly used. Daphne oil is easier to manipulate due to its higher viscosity and lower toxicity. When hydrostaticity at higher pressures is needed, alcohol mixtures are recommended, but special care should be given to the preparation of the pressure cells in order to avoid leakage as these mixtures have a low viscosity.

\section{Scientific examples} \label{Sec:Examples}

  Muon spin rotation/relaxation is been widely applied to magnetic materials due to the high sensitivity of the muon to small
fields and its capability to probe both static and dynamic local
field distributions. ZF ${\mu}$SR is used to investigate microscopic magnetic properties of solids \cite{KuboToyabe_1967,Major,Amato1997,Dalmas}.
It is also valuable for studying materials in which magnetic order is random or of short
range. Moreover, ${\mu}$SR is very helpful to study samples containing multiple phases or
samples which are partially magnetically ordered. This is because muons stop uniformly throughout a sample,
and the amplitudes of the ${\mu}$SR signals arising from the different regions of the sample
are proportional to the volume of the sample occupied by a particular phase.
Thus, ${\mu}$SR provides quantitative information on coexisting and competing
phases in a material.

  The ${\mu}$SR technique provides a powerful tool to measure the important length scales of superconductors, namely the magnetic penetration depth $\lambda$ and the coherence length $\xi$ \cite{Tinkhamsuperconductivity}. A ${\mu}$SR experiment in the vortex state \cite{Abrikosovvortex1} of a type II superconductor allows one to determine ${\lambda}$ in the
bulk of the sample, in contrast to many techniques that probe ${\lambda}$
only near the surface. $\lambda$ is one of the fundamental parameters of a superconductor,
since it is related to the superfluid density $n_{s}$ via 1/${\lambda}^{2}$
= $\mu_{0}$$e^{2}$$n_{s}/m^{*}$ (where $m^{*}$ is the effective
mass of the superconducting carriers). Most importantly, the temperature dependence of ${\lambda}$
is particularly sensitive to the topology of the SC gap.

 In addition to studies of the ideal periodic array of vortices in clean samples,
${\mu}$SR has been shown to be a unique microscopic probe of vortex fluctuations,
pinning, flux-lattice melting, and the decomposition of flux lines into two-dimensional
``pancake" vortices \cite{Leemelting,LeePRL75,AegerterPRB54,AegerterPRB57}.
One can also probe vortices at the surface of a superconductor using ultra-low
energy muons \cite{MorenzoniLEM}.

In the following,a few selected experimental results obtained by performing $\mu$SR experiments under pressure are presented. For further details and examples we refer to the textbooks \cite{Yaouanc_BOOK_2011,Schenck_BOOK_1985}  and original articles \cite{Khasanov84,Bendele2,Khasanov3,Guguchia1,Guguchia2,Shermadini1,Guguchia3,Morenzoni2,Guguchia4,Prando,Maisuradze2,ThedePRL,Egetenmeyer,Ghannadzadeh}.

\subsection{Pressure induced magnetic order in the heavy-fermion URu$_2$Si$_2$}

The heavy-fermion compound URu$_2$Si$_2$ exhibits two successive phase transitions
at 17.5 and 1.4 K. Whereas the transition at 1.4 K signals the occurrence
of unconventional superconductivity, the phase transition at $T_{\rm o}$ = 17.5 K
remains mysterious. Different scenarios have been invoked concerning the true nature of
the phase below $T_{\rm o}$. Models suggest that the observed anomalies are not
related to the magnetic state, but is rather associated with a hidden order parameter.

\begin{figure}[htb]
\centering
\includegraphics[width=0.6\linewidth]{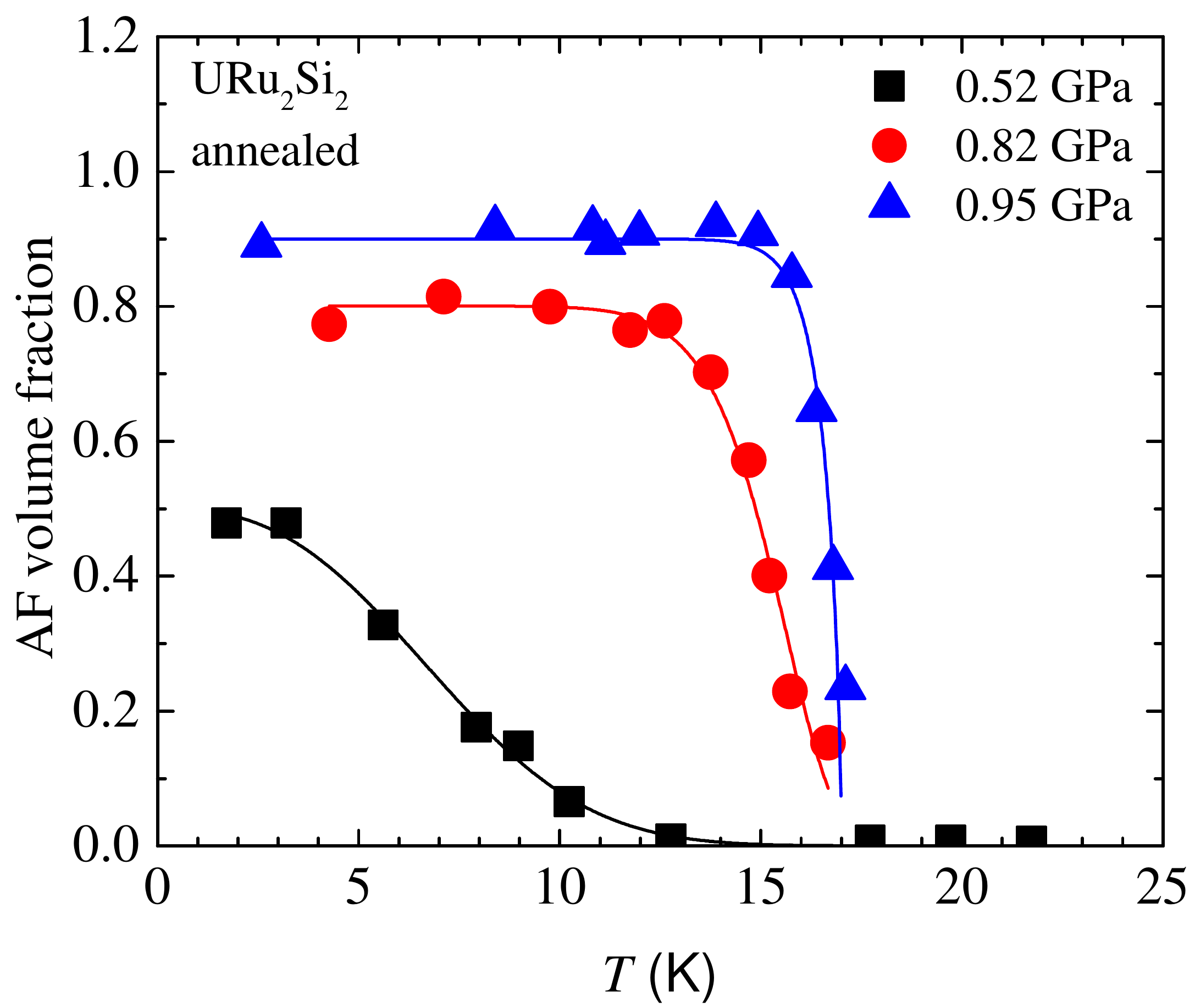}
%
\caption{Temperature and pressure evolution of the antiferromagnetic fraction of URu$_{2}$Si$_{2}$
determined by ${\mu}$SR in an annealed single crystal. For clarity only a few pressure curves are
shown. Note that for low pressures the onset of the magnetic fraction is much lower than the
hidden order temperature $T_{\rm o}$ ${\simeq}$ 17.5 K. (After Ref. \cite{AmatoJPCM}).}
 \label{fig:Phase-diagramm_URuSi}
\end{figure}

In view of its ability to detect different magnetic responses, ${\mu}$SR studies under applied hydrostatic
pressure up to $p$ = 1.5 GPa were performed \cite{Amitsuka2,AmatoJPCM}. The
main finding was the observation of spontaneous muon spin precession in zero applied magnetic
field at pressures above about 0.5 GPa, clearly indicating spontaneous magnetic ordering of the
magnetic moments in the sample. As shown in Fig.~\ref{fig:Phase-diagramm_URuSi}, the magnetic volume fraction is strongly
pressure dependent and starts increasing at temperatures lower than $T_{o}$. A fully
magnetised sample is obtained for pressures of  ${\sim}$ 1 GPa.
Due to the possibility to obtain independently a measure of the magnetic volume and of
the magnetic moment, in this context the ${\mu}$SR data provided a much clearer point of view.
In addition, from the observed behaviour under pressure,
it was demonstrated that the superconducting and magnetic states  are not coupled but rather phase separated with
almost degenerate condensing energies.
%
%


\subsection{Direct observation of the quantum critical point in the heavy fermion CeRhSi$_3$ }

 A drastic and monotonic suppression of the internal fields in the noncentrosymmetric heavy fermion antiferromagnet
CeRhSi$_{3}$  was observed upon increasing the external pressure \cite{Egetenmeyer}. At 2.36~GPa, the ordered magnetic moments are fully suppressed
via a second-order phase transition, and $T_{\rm N}$ is zero, providing direct evidence of the quantum
critical point hidden inside the superconducting phase of CeRhSi$_{3}$.

\subsection{Pressure Induced Static Magnetic Order in Superconducting FeSe$_{1-x}$}



\begin{figure}[b]
\centering
\includegraphics[width=0.8\linewidth]{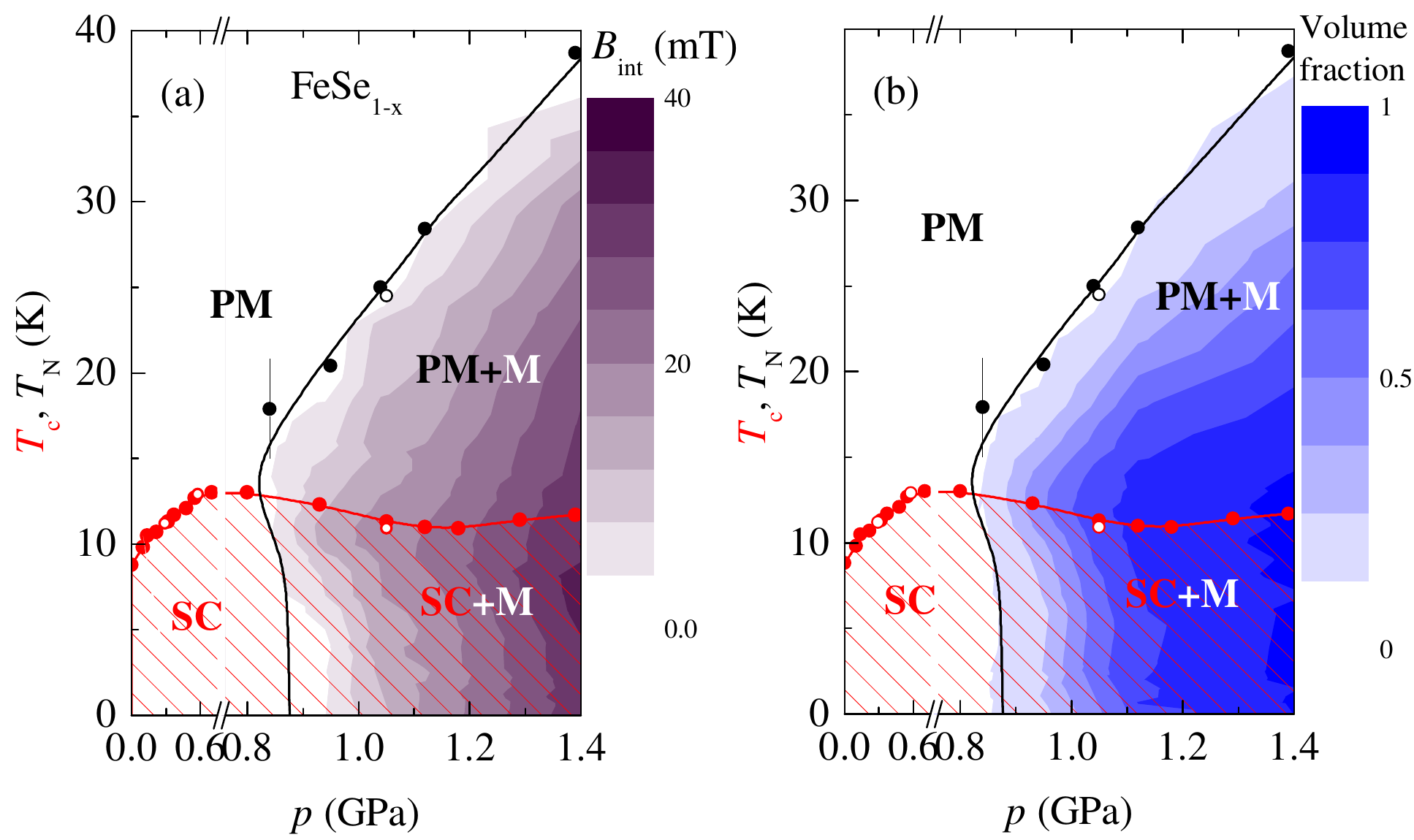}
%
\caption{ (a) Pressure dependence of the superconducting
transition temperature $T_c$, the magnetic ordering temperature
$T_N$, and the internal field $B_{int}$ (magnetic order parameter)
obtained in AC susceptibility and muon-spin rotation experiments.
(b) Pressure dependence of $T_N$, $T_c$, and the magnetic volume
fraction. The $T_c(p)$ and $T_N(p)$ lines are guides for the eye.
The closed and the open symbols refer to FeSe$_{0.94}$ and
FeSe$_{0.98}$ sample. SC, M, and PM denote the superconducting,
magnetic and nonmagnetic (paramagnetic) states of the sample. (After Ref. \cite{BendelePRL}).}
 \label{fig:Phase-diagramm_FeSe}
\end{figure}

A detailed study of the evolution of the superconducting and magnetic properties of FeSe$_{1-x}$  as a
function of pressure and temperature through a combination of AC
susceptibility and muon-spin rotation ($\mu$SR) techniques were recently performed \cite{BendelePRL,Pomjakushina09,Khasanov08_FeSe}.
The obtained phase diagram (see Fig.~\ref{fig:Phase-diagramm_FeSe}) of
FeSe$_{1-x}$ was found to be separated onto three distinct regions.
At low pressures, $0\leq p\lesssim 0.8$~GPa, the samples are
nonmagnetic and $T_c$ increases monotonically with increasing pressure.
In the intermediate pressure region, $0.8\lesssim
p\lesssim1.0$~GPa, $T_c(p)$ decreases with increasing pressure and
the static magnetism develops. In this region of the phase
diagram the superconducting and the magnetic order parameters
coexist and compete on a short length scale. Incommensurate magnetic order
which sets in above $T_c$, becomes partially (or even
fully) suppressed below $T_c(p)$.
At higher pressures, $p\gtrsim1.0$~GPa, $T_c(p)$ shows a tendency
to  rise again. The magnetic order becomes commensurate and both,
the bulk magnetism and the bulk superconductivity coexist within the
whole sample volume.
This exceptional observation provided a new challenge for theories describing the mechanism of high temperature superconductivity.
%

\subsection{Direct evidence for a pressure induced nodal superconducting gap in the Ba$_{0.65}$Rb$_{0.35}$Fe$_{2}$As$_{2}$ superconductor}

   In contrast to other unconventional superconductors,
in the Fe-HTSs both $d$-wave and extended $s$-wave pairing symmetries
are close in energy. Probing the proximity between these very different
superconducting states and identifying experimental parameters that can tune them
is of central interest.

\begin{figure}
\begin{centering}
\includegraphics[width=0.4\columnwidth]{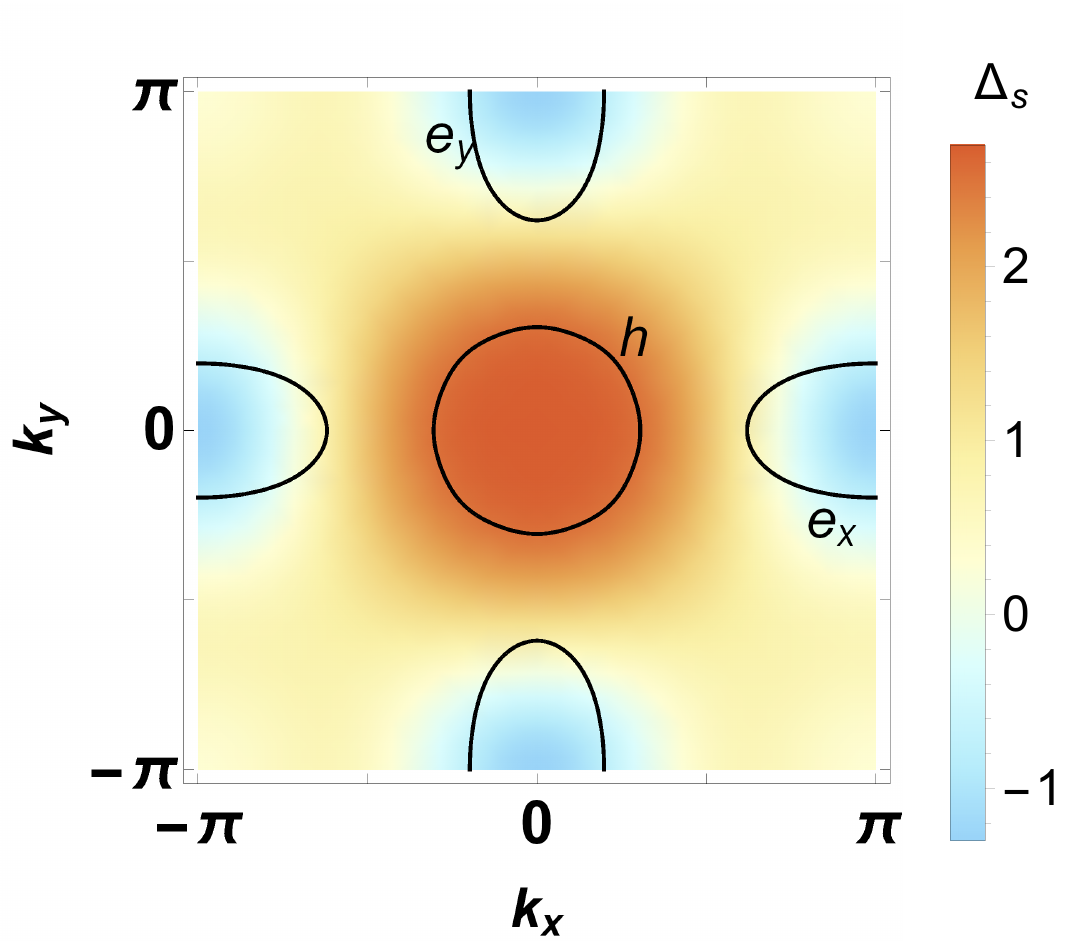}\vspace{-0.2cm}\includegraphics[width=0.4\columnwidth]{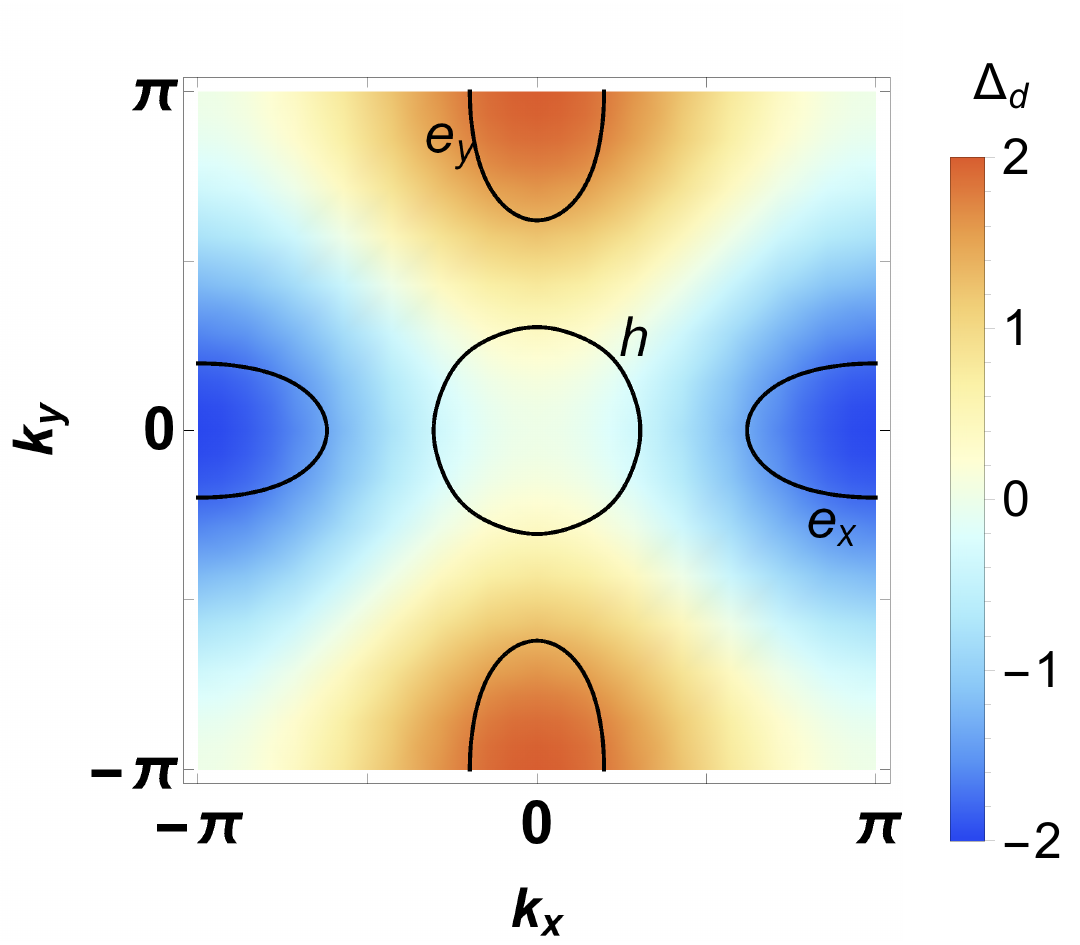}
\vspace{0.1cm}
\par\end{centering}

\caption{Schematic representation of the nodal $s^{+-}$ and $d$-wave
states. In both figures, a color plot of the gap function is superimposed
to a representative Fermi surface consisting of a hole pocket ($h$)
at the center and an electron pocket ($e$) at the borders of the
Brillouin zone. In the nodal $s^{+-}$ states (left panel), the nodes
are not enforced by symmetry (here they are located at the electron
pockets). In the $d$-wave state (right panel), the nodes are enforced
by symmetry to be on the diagonals of the Brillouin zone, and therefore
can only cross the hole pockets. (After Ref. \cite{Guguchia3})}
\label{gap}
\end{figure}

  The zero-temperature magnetic penetration depth ${\lambda}\left(0\right)$
and the temperature dependence of ${\lambda^{\rm -2}}$ were studied in
optimally doped Ba$_{0.65}$Rb$_{0.35}$Fe$_{2}$As$_{2}$ by means
of ${\mu}$SR experiments as a function of pressure up to $p$ ${\simeq}$
2.25 GPa \cite{Guguchia3}. The superconducting transition temperature stays nearly constant under
pressure, whereas a strong reduction of ${\lambda}\left(0\right)$
is observed, possibly related to the presence of a putative quantum
critical point. The main finding is the observation
that pressure promotes a nodal SC gap in Ba$_{0.65}$Rb$_{0.35}$Fe$_{2}$As$_{2}$.
Model calculations favor a $d$-wave over a nodal $s^{+-}$-wave pairing
as the origin for the nodal gap (see Fig.~\ref{gap}).

\subsection{Pressure-induced electronic phase separation of magnetism and superconductivity in CrAs}

At ambient pressure CrAs is characterized by a relatively high N\'{e}el temperature $T_N\simeq 270$~K \cite{Wu_Arxiv_14,Kotegawa_JPSJ_14,Keller_Arxiv_14,Watanabe_69}.
$T_N$  decreases approximately  by a factor of three for pressures ($p$) approaching $\simeq0.7$~GPa, above which the magnetism
completely disappears \cite{Wu_Arxiv_14,Kotegawa_JPSJ_14}. On the other hand superconductivity sets in for
pressures exceeding $\simeq0.4$~GPa thus revealing a range of $0.4\lesssim p \lesssim 0.7$~GPa where superconductivity
and magnetism coexist.

 In view of its ability to provide an additional microscopic point of view of the magnetic and the superconducting properties of CrAs
${\mu}$SR studies under applied hydrostatic pressure were performed \cite{KhasanovSCR}. The obtained results suggest that the pressure-induced transition of CrAs from a magnetic to a superconducting state is characterized by a separation in macroscopic size magnetic and superconducting volumes. The less conductive magnetic phase provides
additional carriers (doping) to the superconducting parts of CrAs. This would naturally explain the substantial increase of both, the transition temperature $T_c$ (from 0.9~K to 1.2~K) and the superfluid density $\rho_s(0)$ (up to $\simeq 150$\%), in the phase coexistence region.

\subsection{Tuning of competing magnetic and superconducting phase volumes in LaFeAsO$_{0.945}$F$_{0.055}$ by hydrostatic pressure}

  It was found that the application of pressure leads to a substantial decrease of
the magnetic ordering temperature $T_{N}$ and a reduction of the magnetic phase volume and, at
the same time, to a strong increase of the superconducting transition temperature $T_{c}$ and the
diamagnetic susceptibility \cite{Khasanov84}. It was also concluded that
the superconducting and the magnetic areas which coexist in the same sample are inclined towards
spatial separation and compete for phase volume as a function of pressure.

\subsection{Superfluid density and superconducting gaps of RbFe$_{2}$As$_{2}$ as a function of hydrostatic pressure}

  Using ${\mu}$SR the ratio of the superfluid density  to the superconducting transition
temperature $T_{\rm c}$ = 2.52(2) K at ambient pressure was determined and found to be much larger in the strongly hole-overdoped
RbFe$_{2}$As$_{2}$ than in high-$T_{c}$ Fe-based and other unconventional superconductors \cite{Shermadini1}.
As a function of pressure $T_{c}$ decreases with a rate of d$T_{\rm c}$/d$p$ = -1.32 K GPa$^{-1}$, {\it i.e.} it is reduced
by 52${\%}$ at $p = 1$~GPa. The temperature dependence of $n_{\rm s}$ is best described by a two gap $s$-wave
model with both superconducting gaps being decreased by hydrostatic pressure until the smaller gap
completely disappears at $p$ = 1 GPa.

\subsection{Strong enhancement of conventional superconductivity near a quantum critical point of (Ca,Sr)$_{3}$Ir$_{4}$Sn$_{13}$ }

\begin{figure}[h]
\centering
\includegraphics[width=0.5\linewidth]{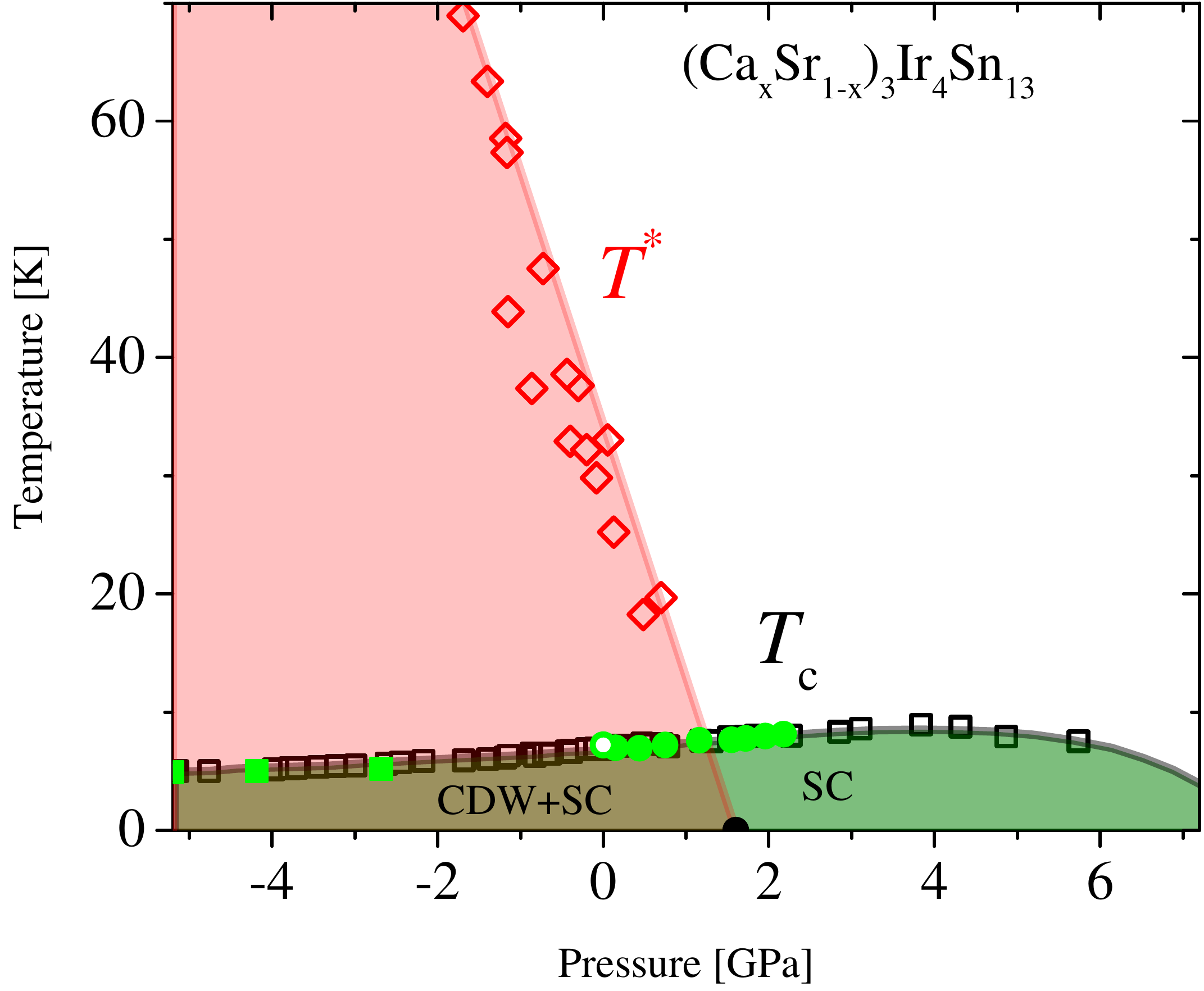}
%
\caption{ Low temperature phase diagram of (Ca$_{1-x}$Sr$_{x}$)$_{3}$Ir$_{4}$Sn$_{13}$  proposed by $\mu$SR investigations of the microscopic
superconducting parameters. They indicate that there are two superconducting phases, one coexisting with CDW and the other
pure. They are separated by a critical point at ${\simeq}$ 1.6 GPa, indicated by a black dot. Green circles and squares: superconducting
transition temperatures of Ca$_{3}$Ir$_{4}$Sn$_{13}$ and Sr$_{3}$Ir$_{4}$Sn$_{13}$ under pressure obtained from the ${\mu}$SR data. (After Ref. \cite{Morenzoni2}).}
 \label{fig:Phase-diagramm_CaSrIrSn}
\end{figure}

  The (Ca$_{1-x}$Sr$_{x}$)$_{3}$Ir$_{4}$Sn$_{13}$ system displays superconductivity and a structural phase transition associated with the formation of a
charge density wave (CDW). ${\mu}$SR studies as a function of pressure revealed a strong enhancement of the superfluid density and the of pairing strength
above a pressure of ${\simeq}$ 1.6 GPa giving direct evidence of the presence
of a quantum critical point separating a superconducting phase coexisting with CDW (see Fig.~\ref{fig:Phase-diagramm_CaSrIrSn}) from a pure
superconducting phase \cite{Morenzoni2}. The superconducting order parameter in both phases has the same $s$-wave
symmetry. Based on the dependence of the effective superfluid density on the critical temperature, it was suggested that
the conventional BCS superconductors in the presence of competing orders or
multi-band structure can also display characteristics of unconventional superconductors.

\subsection{Effect of pressure on the Cu and Pr magnetism in Nd$_{1-x}$Pr$_{x}$Ba$_{2}$Cu$_{3}$O$_{7}$ investigated by muon spin rotation}

A positive pressure effects on the N\'{e}el temperatures of both copper and praseodymium were found for Nd$_{1−x}$Pr$_{x}$Ba$_{2}$Cu$_{3}$O$_{7}$
for the whole range of Pr concentrations (0.3 ${\lesssim}$ $x$ ${\lesssim}$ 1) \cite{Maisuradze_PRB_2013}. These findings clarify some of the puzzles related
to the effect of pressure on superconductivity and magnetism in the praseodymium-substituted hole-doped cuprate systems.

\subsection{Tuning the static spin-stripe phase and superconductivity in La$_{2-x}$Ba$_{x}$CuO$_{4}$ ($x$ = 1/8) by hydrostatic pressure}

 An unusual interplay between static spin-stripe order and bulk
superconductivity was observed in La$_{2-x}$Ba$_{x}$CuO$_{4}$ ($x$ = 1/8) \cite{Guguchia4}.
With increasing pressure the spin order temperature
and the size of the ordered moment are not changing significantly. However, application of
hydrostatic pressure leads to a remarkable decrease of the magnetic volume fraction $V_{m}$(0).
Simultaneously, an increase of the SC volume fraction $V_{sc}$(0) occurs. Furthermore, it was
found that $V_{m}$(0) and $V_{sc}$(0) at all $p$ are linearly correlated: $V_{m}$(0) + $V_{sc}$(0) ${\simeq}$ 1.
These results provide evidence for a competition between bulk superconductivity and static magnetic order in the stripe phase.

\subsection{Muon spin rotation investigation of the pressure effect on the magnetic penetration depth in YBa$_{2}$Cu$_{3}$O$_{x}$}

The pressure dependence of the superfluid density ${\rho}_{{\rm s}}$ as a function of
the superconducting transition temperature $T_{c}$ is found to deviate from the usual Uemura line \cite{Maisuradze2}. The
ratio (${\delta}$$T_{\rm c}$/${\delta}$$P$)/(${\delta}$${\rho}_{{\rm s}}$(0)/${\delta}$$P$) is factor of   ${\sim}$2 smaller than that of the Uemura relation.
In underdoped samples, an increase of the coupling strength with pressure was also found.
In the same work, a special model was also reported for analyzing the ${\mu}$SR spectra of samples with strong magnetic moments in
a pressure cell.

\subsection{Pressure-Induced Quantum Critical and Multicritical Points in a Frustrated Spin Liquid}

 Using high-pressure ${\mu}$SR technique the ($p-T$) phase diagram was established for the quantum spin-liquid compound (C$_{4}$H$_{12}$N$_{2}$)Cu$_{2}$Cl$_{6}$ \cite{ThedePRL}.
At low temperatures, pressure-induced incommensurate magnetic order
is detected beyond a quantum critical point at $p_{\rm c}$ ${\sim}$ 0.43~GPa. An additional phase transition to
a commensurate magnetic order is observed at $p_{1}$ ${\sim}$ 1.34~GPa.
The obtained ($p-T$) phase diagram reveals the corresponding
pressure-induced multicritical point at $p_{1}$, $T_{1}$ = 2.0 K.

\begin{figure}[htb]
\centering
\includegraphics[width=0.5\linewidth]{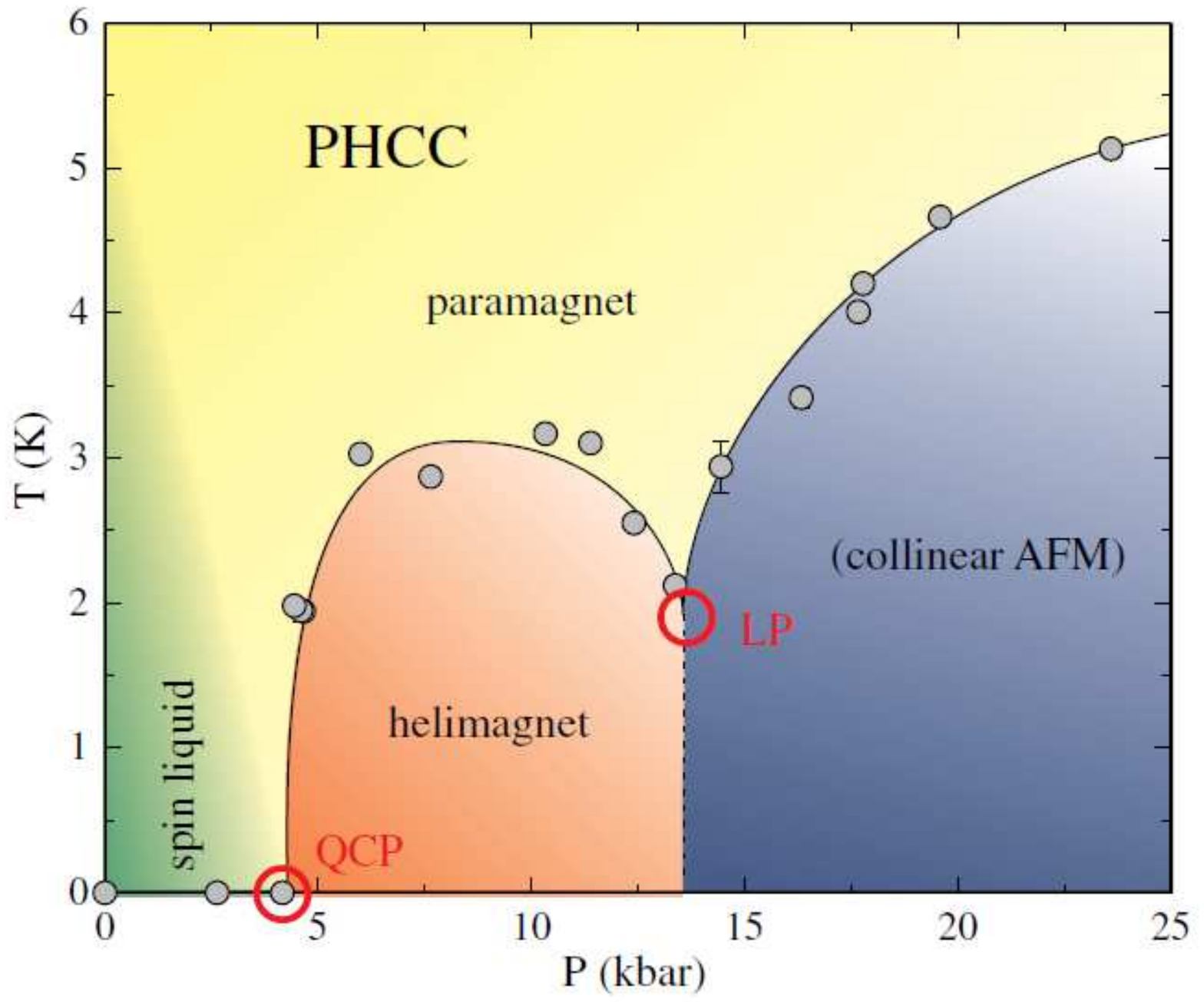}
\vspace{-4.5cm}
\caption{Measured $p-T$ phase diagram of PHCC. Symbols are transition temperatures, as determined by transverse-field
measurements. (After Ref. \cite{ThedePRL}).}
 \label{fig:Phase-diagramm_Thede}
\end{figure}

\subsection{Evolution of magnetic interactions in a pressure-induced Jahn-Teller driven magnetic dimensionality switch}

Pressure induced transition from a quasi-two-dimensional to a quasi-one-dimensional antiferromagnetic phase at 0.91~GPa,
driven by a rotation of the Jahn-Teller axis, was observed in  the coordination polymer CuF$_{2}$(H$_{2}$O)$_{2}$ (pyrazine) \cite{Ghannadzadeh}.
Long-range antiferromagnetic ordering is seen in both regimes, as well as a phase separation in the critical pressure region.
The magnetic dimensionality switching as pressure is increased is accompanied by a halving of the primary magnetic exchange energy $J$
and a fivefold decrease in the ordering temperature $T_{\rm N}$.

\section{Outlook}

In this review paper we have summarized the status of high-presssure research at the Paul Scherrer Institute (Switzerland) by means of muon-spin rotation/relaxaton techniques.
%

Further possible development may be classified along two lines. The first relates to the upgrade of the GPD spectrometer itself. A new detector system  consisting of plastic scintillators read-out by Geiger-mode Avalanche Photodiodes (APD)\cite{Renker_NIM_2006, Stoykov_NIM_2005, Stoykov_PhysB_2009a, Stoykov_PhysB_2009b, Stoykov_JINST_2011, Stoykov_PhProc_2012} is planned to be installed in the next few years. The use of APD's, which are not sensitive to magnetic fields and do not require the usage of long light guides, allows to build a very compact detector system with an optimized detectors geometry thus improving the signal to background ratio.
The second part relates to the improvement of the $\mu$SR pressure cells. This includes: (i) the search for new materials suitable (from the background point of view) for $\mu$SR experiments, (ii) the investigation of new pressure cell designs (as the McWhan-type of cell \cite{Klotz_book_2013}, three wall cells, {\it etc.} ), and (iii) the possible use of anvil pressure cells.  Our preliminary measurements by using Boron-Nitride anvils from te Paris-Edinburgh cell \cite{Klotz_book_2013} reveal that enough muons (up to $\sim 20$~\%) could stop into the sample therefore opening a new era of novel $\mu$SR experiments.

\end{document}